\documentclass[fleqn,usenatbib]{mnras}

\usepackage{newtxtext,newtxmath}
\usepackage[dvipsnames]{xcolor}
\usepackage{soul}

\usepackage[T1]{fontenc}
\usepackage{ae,aecompl}

\usepackage{graphicx}	% Including figure files
\usepackage{lastpage} 
\usepackage{subfig}
\usepackage{multirow}
\usepackage{float}
\usepackage{array,booktabs}
\usepackage{pdflscape}
\newcommand{\GALFORM}[0]{{\tt GALFORM }}
\newcommand{\uncertainty}[3]{${#1}^{\scriptscriptstyle\rm #2}_{\scriptscriptstyle\rm #3}$}

\title{The PAU Survey: The $i$-band galaxy luminosity function from the present-day to $z=2$}

\author[S. Koonkor et al.]{S.~Koonkor,$^{1,2,3}$\thanks{E-mail: suttikoon.koonkor@outlook.com}
C.~M.~Baugh,$^{1,2}$
G. Manzoni,$^{4}$
D. Navarro-Giron\'es,$^{5}$
P. Renard,$^{6,7}$
H. Hoekstra,$^{5}$\newauthor
H. Hildebrandt,$^{8}$
E. Gaztañaga,$^{9,6,7}$
J. Garc\'ia-Bellido,$^{10}$
P. Tallada-Cresp\'i,$^{11,12}$
F. J. Castander,$^{6,7}$\newauthor
J. De Vincente,$^{13}$
R. Casas,$^{6,7}$
R. Miquel,$^{14,15}$
N. Sevilla-Noarbe,$^{13}$
M. Eriksen,$^{14}$
\\ \\
$^{1}$Institute for Computational Cosmology (ICC), Department of Physics, Durham University, South Road, Durham DH1 3LE, UK\\
$^{2}$Institute for Data Science, Durham University, South Road, Durham DH1 3LE, UK\\
$^{3}$National Astronomical Research Institute of Thailand, Don Kaeo, Mae Rim, Chiang Mai, 50180, Thailand\\
$^{4}$State Key Laboratory of Displays and Opto-Electronics, The Hong Kong University of Science and Technology, Hong Kong, China\\  
$^{5}$Leiden Observatory, Leiden University, Einsteinweg 55, 2333 CC Leiden, the Netherlands\\
$^{6}$Institute of Space Sciences (ICE, CSIC), Campus UAB, Carrer de Can Magrans, s/n, 08193 Barcelona, Spain\\
$^{7}$Institut d'Estudis Espacials de Catalunya (IEEC), E-08860 Castelldefels (Barcelona), Spain\\
$^{8}$Ruhr University Bochum, Faculty of Physics and Astronomy, Astronomical Institute (AIRUB), German Centre for Cosmological Lensing, 44780 Bochum, \\
Germany \\
$^{9}$Institute of Cosmology \& Gravitation, University of Portsmouth, Dennis Sciama Building, Burnaby Road, Portsmouth PO1 3FX, UK\\
$^{10}$Instituto de F\'isica Te\'orica CSIC/UAM, Universidad Aut\'onoma de Madrid, 28049 Madrid, Spain \\
$^{11}$Centro de Investigaciones Energéticas, Medioambientales y Tecnológicas (CIEMAT), Avenida Complutense 40, 28040 Madrid, Spain.\\
$^{12}$Port d’Informaci\'o Cient\'ifica (PIC), Campus UAB, C. Albareda s/n, 08193 Bellaterra (Barcelona), Spain.\\
$^{13}$Centro de Investigaciones Energéticas, Medioambientales y Tecnológicas (CIEMAT), Avenida Complutense 40, 28040 Madrid, Spain.\\
$^{14}$Institut de F\'isica d’Altes Energies (IFAE), The Barcelona Institute of Science and Technology, Campus UAB, 08193 Bellaterra (Barcelona), Spain.\\
$^{15}$Instituci\'o Catalana de Recerca i Estudis Avan\c{c}ats (ICREA), 08010 Barcelona, Spain.\\
}

\date{Accepted XXX. Received YYY; in original form ZZZ}

\pubyear{2025}

\begin{document}
\label{firstpage}
\pagerange{\pageref{firstpage}--\pageref{lastpage}}
\maketitle

%%%%%%%%%%%%%%%%%%%%%%%%%%%%%%%%%%%%%%%%%%%%%%%%%%
%%%%%%%%%%%%%%%%%%% ABSTRACT %%%%%%%%%%%%%%%%%%%%%
%%%%%%%%%%%%%%%%%%%%%%%%%%%%%%%%%%%%%%%%%%%%%%%%%%
\begin{abstract}
We present a measurement of the $i$-band galaxy luminosity function from the present-day to $z = 2$, using over 1.1 million galaxies from the Physics of the Accelerating Universe Survey (PAUS). PAUS combines broad-band imaging from the Canada-France-Hawaii Telescope Lensing Survey with narrow-band photometry from PAUCam, enabling high-precision photometric redshifts with an accuracy of $\sigma_{68} (\Delta z) = 0.019$ down to $i_{\textrm{AB}} = 23$. A synthetic lightcone mock catalogue built using the \texttt{GALFORM} semi-analytic model is used to simulate PAUS selection effects and photometric uncertainties, and to derive a machine-learning based estimate of the $k$-correction. We recover rest-frame $i$-band luminosities using a random forest regressor trained on simulated $ugriz$ photometry and redshifts. Luminosity functions are estimated using the $1/V_{\textrm{max}}$ method, accounting for photometric redshift and magnitude errors, and validated against mock data. We find good agreement between observations and models at $z < 1$, with increasing discrepancies at higher redshifts due to photometric redshift outliers. The bright-end of the luminosity function becomes flatter at high redshift, primarily driven by redshift errors. We show that the faint-end of the luminosity function becomes more incomplete with increasing redshift, but is still useful for constraining models. We analyze the red and blue galaxy populations separately, observing distinct evolutionary trends. The model overpredicts the number of both faint red and blue galaxies. Our  study highlights the importance of accurate redshift estimation and selection modeling for robust luminosity function recovery, and demonstrates that PAUS can characterise the galaxy population  with photometric redshifts across a wide redshift baseline.
\end{abstract}


%%%%%%%%%%%%%%%%%%%%%%%%%%%%%%%%%%%%%%%%%%%%%%%%%%
%%%%%%%%%%%%%%%%% BODY OF PAPER %%%%%%%%%%%%%%%%%%
%%%%%%%%%%%%%%%%%%%%%%%%%%%%%%%%%%%%%%%%%%%%%%%%%%

%\tableofcontents
%\clearpage

\section{Introduction}

The luminosity function provides a basic characterisation of the galaxy distribution. The location of the break in the luminosity function and the slope of the faint end encode information about the strengths of various processes such as the cooling of gas, the rate of star formation and feedback from supernovae, and the heating of gas by active galactic nuclei \citep{Efstathiou00, Efstathiou03, Benson03}. 
Here we revisit the seminal estimation of the evolution of the galaxy luminosity function by \cite{Lilly1995}, using a modern survey with almost $1\,500$ times more galaxies, guided by a synthetic mock catalogue that allows us to quantify systematic effects.

\cite{Lilly1995} measured the evolution of the galaxy luminosity function using imaging in 5 fields each of $10^{\prime} \times 10^{\prime}$ and spectroscopy to obtain 730 galaxy redshifts. To make as complete a census of the galaxy population as possible it is important to mitigate any biases in the sample selection. For example, placing a requirement on measuring spectroscopic redshifts with high confidence may result in a bias towards galaxies with strong emission lines. Also, the relative expense of spectroscopy compared to photometry may lead to smaller solid angles being covered, causing an under or over-representation of galaxy clusters. These selection effects can be avoided by using photometric redshifts over a larger solid angle.

%The galaxy luminosity function has been estimated from many surveys over more than 60 years. The estimate by \citet{Lilly1995} is notable for studying the evolution of the overall luminosity function and that of red and blue galaxies over a substantial redshift interval using a single survey. This simplifies the analysis and makes the comparison between the luminosity functions at different redshifts easier. Lilly et~al. studied the luminosity function of red and blue galaxies measured using the Canada-France Redshift Survey with 731 galaxies out to redshift of $z \sim 1$. As all of these galaxies had spectroscopic redshifts, this may introduce a bias due to the redshift completeness of star forming galaxies with strong emission lines could be higher. 
  
A series of photometric and spectroscopic surveys have built upon the early work by \citet{Lilly1995}. 
The Classifying Objects by Medium-Band (MB) Observation in 17 Filters (COMBO-17 Survey: \citealt{COMBO17_2003}) measured the luminosity function from a sample of $\sim 25\,000$ galaxies using photometric redshifts derived from 12 MB filters (with a full width half maximum (FWHM) varying from 140 to 310~\AA \, and $UBVRI$ broad-band (BB) filters covering a total area of $0.78$~deg$^2$. The photometric redshift error was $\sigma_{68}(\Delta_z) = 0.03$ to $R=24$. Similarly, the Advanced Large Homogeneous Area Medium-Band Redshift Astronomical (ALHAMBRA survey: \citealt{ALHAMBRA2008,  Molino+2014}) measured $\sim 438\,000$ galaxy redshifts using 20 MB optical filters (with a FWHM of 310~\AA) over a total solid angle of 2.79~deg$^2$ and achieved $\sigma_{68}(\Delta z) = 0.01$ for $i_{\rm{AB}} < 22.5$. Despite the improvement in the luminosity function estimates from these larger samples which results in a reduced sensitivity to the effects of large scale structure, the photometric redshifts are still substantially less accurate than typical spectroscopic estimates. This can lead to the evolution in the measured luminosity function being mis-estimated, when photometric redshift outliers are assigned to the wrong redshift bin in the analysis. 

Large spectroscopic surveys have provided high-precision measurements of the local luminosity function, including the 2-degree galaxy redshift survey \citep{Norberg2002} and the main Sloan Digital Sky Survey \citep{Blanton2003}, both with median redshifts around $z \sim 0.1$. The deeper Galaxy and Mass Assembly (GAMA) survey measured the luminosity function in several BB filters out to $z \sim 0.5$ \citep{Loveday2012}. The Dark Energy Spectroscopic Instrument (DESI) Bright Galaxy Survey reaches a similar depth to GAMA but over a much larger solid angle, pushing the luminosity function estimates out to $z \sim 0.6$ \citep{Moore+25}. Other surveys have probed intermediate redshifts, but typically adopt a colour selection to target a particular redshift interval without covering $z=0$ (e.g. \citealt{Davidzon2013}). Some spectroscopic surveys have sampled a wide baseline in redshift ($0<z<2$) but at the expense of covering a relatively small solid angle, of order one square degree \citep{Ilbert2005}.

The Physics of the Accelerating Universe Survey (PAUS: \citealt{Eriksen2019,Padilla2019}) is the natural extension to the imaging surveys mentioned above, with narrower filters, so it can produce improved photometric redshift measurements. PAUS has achieved photometric redshift errors that are less than a factor of ten larger than the root mean square (RMS) of spectroscopic redshifts\footnote{For reference, the target random measurement error in redshift for the Dark Energy Spectroscopic Instrument is $\sigma_{z}/(1+z) = 1.4 \times 10^{-4}$ \citep{DESISV}.} rather than the two orders of magnitude larger errors typically achieved with a handful of BB filters. PAUS also covers a larger solid angle on the sky than previous imaging surveys which used medium- and narrow-band (NB) filters (\citealt{Alarcon+21, Navarro2024}).  

Here, we analyse the PAUS observations alongside a mock catalogue provided by \cite{Manzoni2024}. This allows us to assess the estimation of the galaxy luminosity function for systematic effects introduced by the survey selection. In particular, when we apply the same survey selection effect as applied to the observations, the mock will allow us to see the range of luminosities over which the estimate of the luminosity function becomes incomplete. This occurs because we estimate the luminosity function in the selection band; with increasing redshift, the observed selection band moves to shorter wavelengths in the rest frame, which means that some galaxies are missed at faint magnitudes because of their colour. Here we do not attempt to correct for this incompleteness, but argue that it is still useful to use the measured luminosity function where it is incomplete to constrain galaxy formation models. We also use the mock to devise a $k$-correction that takes observable quantities as input.

The layout of this paper is as follows. 
Section 2 describes the observational dataset  (\S~2.1) and the mock catalogue (\S~2.2). The modelling of errors in the galaxy photometry and photometric redshifts is described in \S~2.3. The modelling of the $k$-corrections in the broad-bands, and the estimation of rest-frame absolute magnitude is presented in \S~2.4. In Section 3 we present the results for the estimation of the luminosity function using the mock catalogue, in which case we know the value of the underlying luminosity function. Section 4 presents the results for the luminosity function estimate from the PAU survey. Section 5 discusses the effect of \texttt{GALFORM} parameter variations on the luminosity function. Our conclusions and the discussion are presented in Section 6.

\section{The datasets: observations and mock catalogue}

Here we give a brief description of the observational data used, the PAU Survey (\S~2.1), and the mock catalogue (\S~2.2). The errors in the galaxy photometry and photometric redshifts applied to the galaxies in the mock are described in \S~2.3. The modelling of the $k$-corrections in the broad-bands, and the estimation of rest-frame absolute magnitude is presented in \S~2.4.

\subsection{The Physics of the Accelerating Universe Survey (PAUS)}
The Physics of the Accelerating Universe Survey (PAUS) is a narrow band imaging survey that used the PAUCam instrument \citep{Padilla2019}, mounted on the 4.2-m William Herschel Telescope. PAUS took photometry in 40 NB filters. Each NB has a FWHM of 130~\AA \ with centres spaced by 100 \AA \ over the wavelength range of 4500--8500~\AA. PAUS started collecting data in 2015 in four different fields, including the COSMOS field \citep{Scoville+07}, the Canada-France-Hawaii Telescope Legacy Survey (CFHTLS) W1 and W3 fields \citep{Cuillandre+12, Erben2013}, and the KiDS/GAMA G09 field \citep{deJong+13}. The COSMOS field covers 2~deg$^2$ and has been targeted as a calibration field because it has been covered many times with photometric and spectroscopic surveys with a combination of relatively high completeness and depth. Instead we use the much larger wide fields, focusing on the W1 and W3 fields, which together have 31.71~deg$^2$ of PAUS NB imaging processed and available. We consider the sample to a depth of $i_{\rm AB}=23$, yielding more than $\sim$ 1.1 million galaxies (see Table \ref{Tab:PAUS_areas_n_gal}). 

PAUS NB fluxes are measured by using forced photometry of objects previously detected in the Canada-France-Hawaii Telescope Lensing Survey (CFTHLenS: \citealt{Erben2013}) and the Kilo Degree Survey (KiDS: \citealt{Kuijken+15}); see \cite{Serrano2023} for a description of the NB photometry.  The improved spectral resolution of the PAUS NB photometry allows more accurate photometric redshift estimates, reaching an accuracy of $\sigma_{68}(\Delta z)/(1+z) \approx 0.003$ and $0.009$ for galaxies brighter than magnitude $i_{\rm AB} \sim 18$ and $i_{\rm AB} \sim 23$,  respectively for the COSMOS field (\citealt{Alarcon+21}; see also \citealt{Eriksen2020,Soo2021}).
For the wide fields, PAUS obtains $\sigma_{68}(\Delta_z)/(1+z) \approx 0.019$ for all galaxies with $i_{AB} < 23$ (\citealt{Navarro2024,Daza2024}). 
The photometric redshifts measured with PAU narrow bands are approximately 10 times better than those obtained when using a small number of broad bands (see \citealt{Eriksen2019}).

\begin{table}
\centering
\begin{tabular}{ cccc } 
\hline
Fields & W1 & W3 & total\\
\hline
Area (${\rm deg}^2$)&  11.18 & 20.53 & 31.71\\ 
\# Galaxies& 387\,051 & 738\,884 & 1\,125\,935 \\ 
\hline
\end{tabular}
\caption{The solid angle covered (in square degrees) by PAUS in the CFHTLenS W1 and W3 fields (row two) and the number of galaxies with PAUS NB photometry to a depth of $i_{\rm AB}=23$ (row three).}
\label{Tab:PAUS_areas_n_gal}
\end{table}

\subsection{The lightcone mock catalogue}

We use the mock PAUS catalogue built by \cite{Manzoni2024}, who followed the steps set out in \cite{Merson2013} and \cite{Stothert+18}. 
Manzoni et~al. combined an $N$-body simulation that follows the growth of structure in dark matter, with a treatment of the physics of the baryonic component, made using the \GALFORM semi-analytical model of galaxy formation
\citep{Cole2000,Lacey2016}. \GALFORM predicts the number of galaxies in each dark matter halo, along with their intrinsic properties such as their stellar mass, star formation rate, and luminosity in different bands. 

%to make an $ab$ $initio$ calculation of the formation and evolution of galaxies during the hierarchical growth of structure in the dark matter. \GALFORM \ models the physics of the following processes; (i) the formation and merging of dark matter haloes; (ii) the formation of galactic disks from the shock heating and radiative cooling of gas inside dark matter haloes; (iii) star formation in galaxy disks and in starbursts; (iv) feedback from supernovae, from active galactic nucleus (AGN), and from photo-ionization of the intergalactic medium (IGM); (v) galaxy mergers driven by dynamical friction and bar instabilities in galaxy disks; (vi) calculation of the sizes of disks and spheroids, and (vii) chemical enrichment of stars and gas.

%The spatial distribution of galaxies in the mock catalogue is calculated by considering the observer's past lightcone. The galaxy formation model is implemented in an N-body simulation to perform this step. The dark matter halo merger trees of the galaxy formation model are extracted from the N-body simulation (see Merson et al. 2013; Jiang et al. 2014). 
\citet{Manzoni2024} used the \cite{Lacey2016} version of \GALFORM in which the parameters related to galaxy formation processes have been re-calibrated following the implementation of the model into the Planck Millennium $N$-body simulation (here after PMILL, \citealt{Baugh2019}). The PMILL has a mass resolution that is almost 10 times better than that in previous $N$-body simulations used with \GALFORM and has the total number of different redshift snapshots of 271 (compared to the 60 typically available).  The simulation box is effectively replicated in space to cover the requested solid angle (here set to 100~deg$^2$) and redshift interval ($0<z<2$). The position and properties of a galaxy are determined by the redshift at which it crosses the observer's lightcone. The two simulation snapshots on either side of the lightcone crossing redshift are used to interpolate the galaxy position and observed magnitude\footnote{The observed magnitude filter depends on the redshift at which the galaxy is observed. \citet{Merson2013} showed that this interpolation approach removed any indication of the location of the simulation outputs when plotting the observed colour against redshift. \cite{Manzoni2024} showed that there is no trace of the simulation snapshots in the photometric redshifts recovered from the mocks built using the interpolated observed magnitudes.}; the high cadence of outputs in the PMILL makes this process more accurate compared with previous simulations (e.g. that available to \citealt{Stothert+18}).

\begin{figure}
\centering
    {\includegraphics[width=0.5 \textwidth]{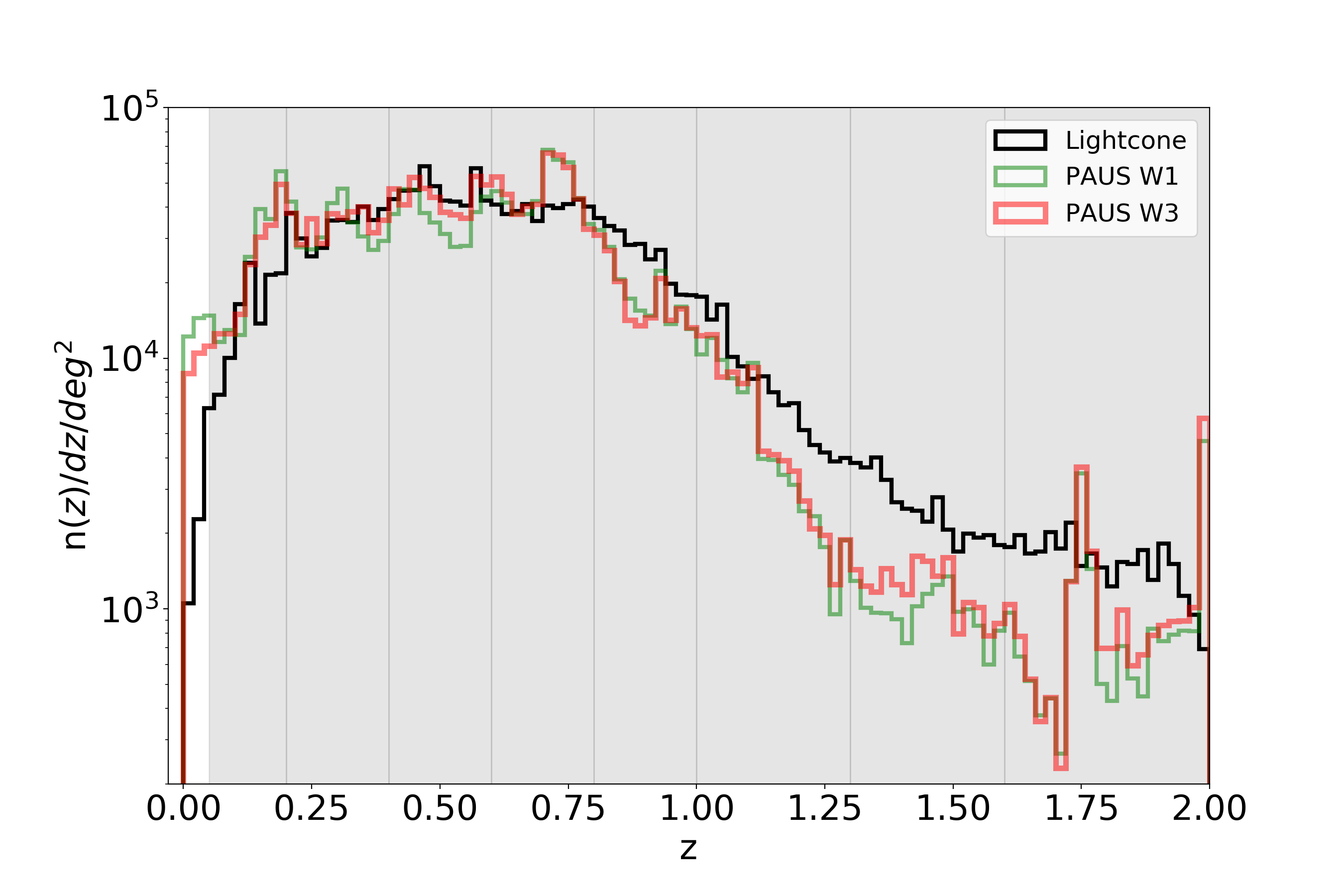}}%
\caption{ Redshift distribution (normalised to the number of galaxies per unit redshift and per square degree) of the \texttt{GALFORM} lightcone mock catalogue (black), PAUS W1 (green), and PAUS W3 (red) to $i_{\rm AB} = 23$. The vertical grey bands show the redshift bins used for calculating the luminosity function.}%
    \label{fig:redshift_distribution}
\end{figure}

The basic properties of the mock, such as the number counts of galaxies, their redshift distribution, and the evolution of the distribution of observed colour were presented in \citet{Manzoni2024}  and found to be in good agreement with the observations from PAUS. Fig.~\ref{fig:redshift_distribution} shows the redshift distribution of galaxies obtained from the mock catalogue of \citet{Manzoni2024} compared to the observation from the PAUS survey in the W1 and W3 fields. The black line represents the "true" spectroscopic redshift distribution of the mock catalogue and the photometric redshift distributions from the PAUS W1 and W3 fields are shown as the green and red lines, respectively. The bin-to-bin fluctuations in the black curve are smaller than those seen for the W1 and W3 fields as the full mock corresponds to a much larger solid angle than either of the W fields. Moreover, the mock and the observation agree better when taking observational uncertainties into account (see \S~\ref{sec:photoz_error}).

\subsection{Including errors in photometry and photometric redshift estimation}
\label{sec:photoz_error}

We also consider the effect of errors in the galaxy photometry and photometric redshift estimation on the recovered luminosity function, by perturbing these quantities in the Manzoni et~al. mock catalogue.

Photometry errors are assumed to have a Gaussian distribution in magnitude. The perturbed magnitude in the band labeled by $j$, $m_{j}^{\rm obs}$ is obtained by adding a Gaussian distributed quantity, $x$, which has zero mean and a variance of ${\sigma}_{j}$ to the true magnitude predicted by \texttt{GALFORM}, $m_{j}^{\rm true}$:

\begin{equation}
    m_{j}^{\rm obs} = m_{j}^{\rm true} + x.
\end{equation}

The variance of the Gaussian is related to the signal-to-noise ratio in band $j$, $(S/N)_{j}$ by:

\begin{equation}
    {\sigma}_{j} = \frac{2.5}{\ln 10}\frac{1}{(S/N)_{j}}.
\end{equation}
 The signal-to-noise ratio is set to 5 at the magnitude limit in a given band. The broad band and narrow band magnitude limits for the PAUS mock are given in \citet{Manzoni2024}. This model assumes that all galaxies are treated as point sources.

%Manzoni et~al. showed that these photometry errors led to similar photometric redshift errors to those estimated for the PAUS data when fed through a photometric redshift code. 

\begin{figure}
\centering
    {\includegraphics[width=0.47\textwidth]{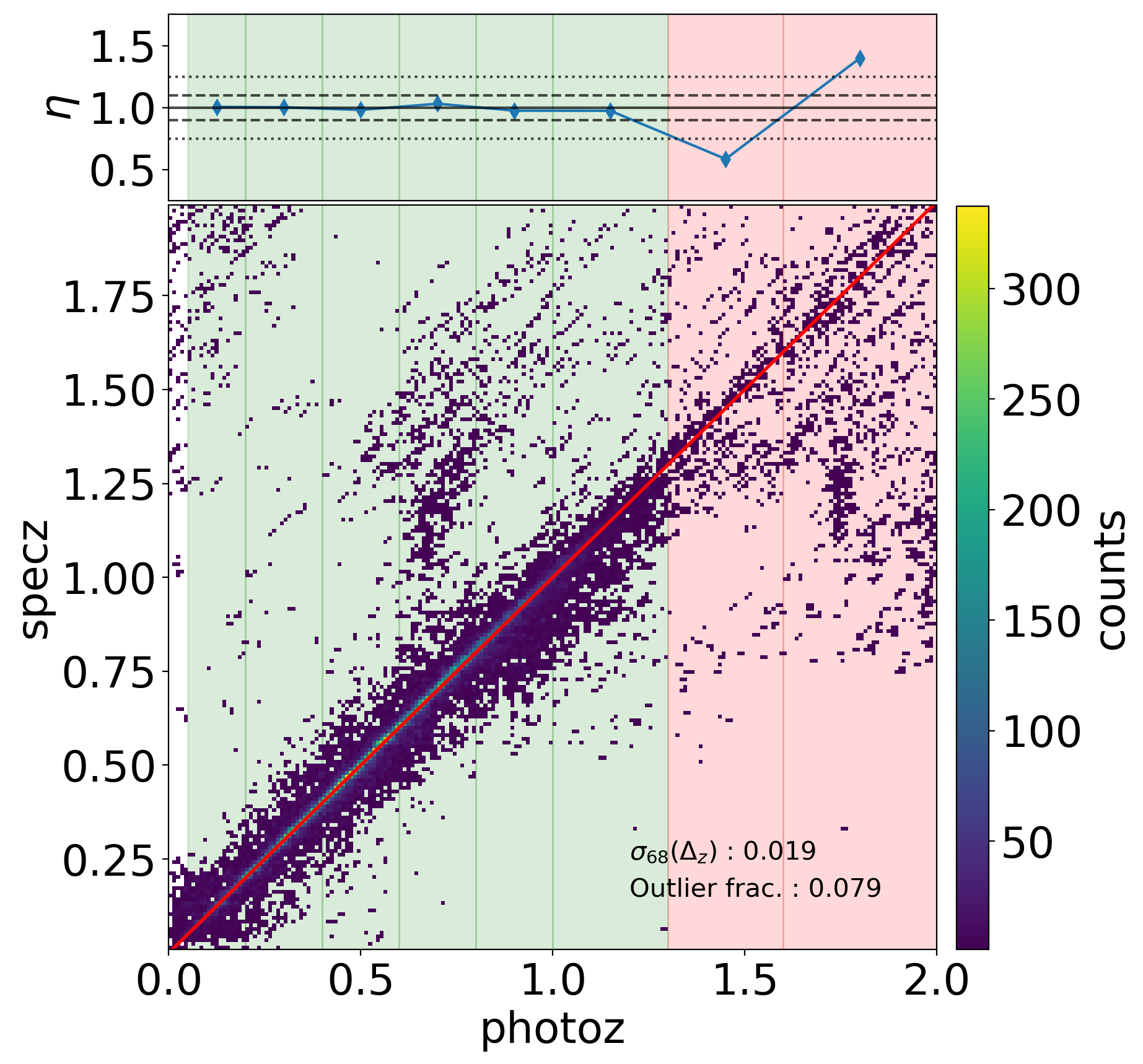}}%
\caption{The photometric redshift estimated from the random sample of galaxies drawn from the lightcone mock catalogue to $i_{\rm AB}=23$ using the \texttt{BCNZ}2 code. The bottom panel shows the true (specz) redshifts from the mock catalogue on the $y$-axis compared to the \texttt{BCNZ}2 estimated redshifts on the $x$-axis. The top panel shows the parameter $\eta$, which is the number of galaxies with photometric redshifts in a particular redshift bin (indicated by the shaded vertical bands) divided by the original number of galaxies in the bin (i.e. using the spectroscopic or true redshift value). The horizontal dashed and dotted lines in the upper panel indicate errors on this number ratio of $10$ per cent and $25$ per cent, respectively. The green and red regions show the redshift bins with $|\eta - 1|$ lower and greater than 10 per cent, respectively}%
\label{fig:specz_photoz_BCNz}%
\end{figure}

The photometric redshift code \texttt{BCNZ}2 \citep{Eriksen2019} was run using a random selection of galaxies from the Manzoni et~al. catalogue, to reduce the computational overhead. This exercise gives the distribution of estimated photometric redshifts as a function of the true redshift, including outliers. We sample this distribution in narrow running bins of the true redshift, to estimate the photometric redshifts for all galaxies in the mock. Using this approach, we have tested that we can recover the scatter and outlier fraction obtained when running the photometric redshift code directly as shown in Fig.~\ref{fig:specz_photoz_BCNz}. Note that this exercise does not use the emission lines predicted by \texttt{GALFORM} (see \citealt{Baugh2022}), so the performance of the photometric redshift estimator is somewhat poorer than it would be for the real data, thus giving a conservative estimate of the errors. Our approach of sampling the error distribution rather than trying to model it using e.g. a Gaussian, is vindicated by the errors above $z \sim 1$. At these redshifts, the fraction of outliers increases substantially, to the extent that it is hard to define the scatter using the central 68 per cent of galaxies (because an increasing fraction of the central 68 percentile range of estimated redshifts corresponds to outliers). In these high redshift bins, the photometric redshift errors can lead to a distortion in the number of galaxies in the redshift bins used to estimate the luminosity function. The actual number of galaxies in these bins, shaded pink in Fig.~\ref{fig:specz_photoz_BCNz}, departs from the expected number (calculated using the cosmological or spectroscopic redshifts) by more than a factor of $10$ per cent.

\subsection{$k$- correction and estimation of the rest-frame magnitude}

The luminosity of a galaxy is an intrinsic property that is proportional to the number of stars in the galaxy, their age, and metallicity (see, for example, the review by \citealt{Conroy2013}). When a galaxy survey covers a substantial baseline in redshift, a given observed filter samples progressively shorter wavelengths in the rest frame of a galaxy spectral energy distribution as the redshift increases. For simplicity, it is common practice to compare the galaxy luminosity measured at different redshifts at a fixed wavelength. We select galaxy samples in an observed band, by applying the $i$-band apparent magnitude limit in our case, and typically estimate the luminosity function at a fixed reference redshift, $z=0$, also in the $i$-band.
%However, the luminosity of a galaxy can change as new stars form and evolve. When measuring the galaxy luminosity function, it requires one to know the intrinsic luminosity of galaxies from observed fluxes. Hence, it is necessary to correct for the fact that a fixed observed bandpass corresponds to a different range of wavelengths in the rest frames of galaxies at different redshifts, 
The adjustment between the observed $i$-band for a galaxy at redshift $z$ and the $z=0$ rest-frame $i$-band is called the $k$-correction \citep{Hogg_Kcorrection, Loveday+12}. 

The luminosity of a galaxy can be $k$-corrected if the shape of the galaxy's spectral energy distribution (SED) and its redshift are known. One can obtain the galaxy SED by using a code that fits empirical SED templates to the observed photometry or which generates synthetic SEDs by assuming a star formation history (and more rarely, a chemical evolution history; e.g. \citealt{prospector2021}). However, obtaining the SED for every single galaxy in a survey and its associated mock catalogue (which could cover a larger solid angle)  could be computationally expensive, especially for a wide and deep survey like PAUS.

\GALFORM predicts the star formation history and chemical evolution of the disk and bulge components of model galaxies. A model is applied to calculate the attenuation of starlight by dust (see \citealt{Lacey2016} for details). A  rest frame and observed frame magnitude can be predicted for each galaxy, so we know the \textit{exact} $k$-correction for each galaxy at the redshift of its lightcone crossing. 

%As we describe in the previous section, a physically motivated model using of the \GALFORM semi-analytical model of galaxy formation on the merger histories of the dark matter haloes from the Planck Millennium allows the model to predict the luminosity or magnitude of the model galaxies. 

%The mock catalogue provides the rest-frame absolute magnitude for every object and it also calculates the observer-frame apparent magnitude using the band-shifting technique of a desired band pass on the galaxy's SED mimicking the real observation. With information we obtain from the mock catalogue, we know exactly what the $k$-correction for every galaxy in a desired redshift range. 

\begin{figure}
\centering
    {\includegraphics[width=0.47\textwidth]{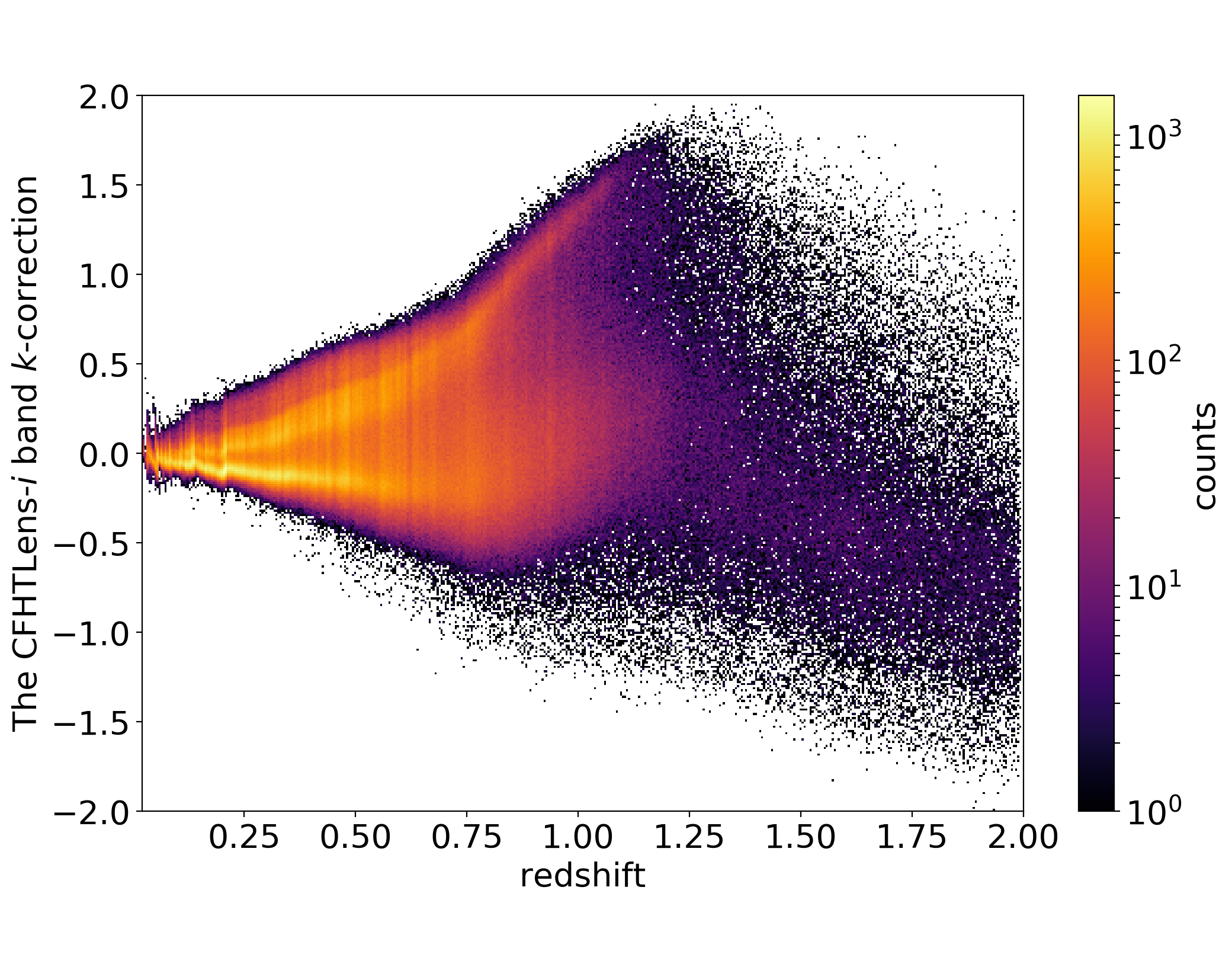}}%
\caption{The exact $i-$band $k$-correction as a function of redshift predicted by \texttt{GALFORM} from the lightcone mock catalogue. The pixel shading shows the galaxy number count for a sample with $i_{\rm AB}=23$, with black representing 1 galaxy and the brightest colour pixel having $\sim1\,300$ galaxies as shown in the colour bar on the right.}%
    \label{fig:GALFORM_kcorrection}%
\end{figure}

Fig.~\ref{fig:GALFORM_kcorrection} shows the distribution of the $k$-correction predicted by \texttt{GALFORM} as a function of redshift for every galaxy in the lightcone mock catalogue built for PAUS to $i_{\rm AB}=23$. The distribution is clearly bimodal with two distinct populations apparent, corresponding to red and blue galaxies (see the plot of observed $g-r$ colour versus redshift in \citealt{Manzoni2024}). The ``positive'' and ``negative'' branches of the $k$-corrections in Fig.~\ref{fig:GALFORM_kcorrection} corresponds to red and blue galaxies, respectively. 
%The two populations arise from the fact that the $k$-correction value depends on the type of the galaxy as we expect  \citep{Manzoni2024}. 

A simple model for the $k$-correction would be to fit a parametric function of redshift to the corrections predicted for red and blue galaxies, tracing the ridges in the galaxy distribution in Fig.~\ref{fig:GALFORM_kcorrection}. Although the peaks in the $k$-corrections are well defined, there is a considerable scatter around the ridges, so such a model would lead to significant errors in the $k$-correction, and in turn in the rest frame $i$-band absolute magnitude. Errors in the absolute magnitude can affect the shape of the estimated luminosity function, particularly at the bright end.  

To reduce this problem, the $k$-correction is often calculated using multiple colour classes (e.g. \citealt{Tamsyn2014}). This approach works by employing the parametric $k$-correction of the colour class that is closest to the colour of a given galaxy. Note this approach does not include any correction due to the formation of new stars over the redshift interval on which the $k$-correction is being applied or the dimming of the existing stars (the evolutionary correction to the luminosity). So whilst this approach has the advantage of providing a $k(z)$ curve, it is not clear that a given galaxy would remain within the colour class that it is assigned to at redshift $z$. 

Note that sometimes an evolutionary correction is also applied to the rest frame luminosity, to account for the change in the stellar population from the redshift of lightcone crossing to $z=0$ due to ongoing star formation in the galaxy, as well as the ageing of the existing stellar population. We do not attempt to make any such correction. Instead, differences in the luminosity function between redshifts reflect this evolution, as well as changes in number density at a given luminosity due to galaxy mergers.

\subsubsection{$k$-corrections using machine learning}
%Even though Fig.~\ref{fig:GALFORM_kcorrection} shows that the $k$-correction distribution may show two main population of galaxies, one may assign a $k$-correction value to a galaxy based on its color. With this approach, we find that blue galaxies have a broad scatter compared to red galaxies which gives a large rest-frame absolute magnitude prediction errors. The magnitude error is crucial for the luminosity function measurement, especially at the bright-end due to the Eddington bias. 

Rather than using a single observed broad-band colour to predict the galaxy $k$-correction, we use all of the available photometry in the $u g r i z$ filters. We use machine learning to find the relation between the $ugriz$ photometry, the redshift and the $k$-correction predicted by \texttt{GALFORM}. The method we chose is the random forest regression algorithm from the publicly available package \texttt{RandomForestRegressor} (RFR) from Scikit-Learn \citep{ScikitLearn2011}. 

RFR is computationally efficient and its performance is not overly sensitive to the choice of hyper-parameter values. Nevertheless, to find the best RFR model for our problem, we selected the hyper-parameter values using the random grid search technique combined with $k$-fold cross-validation to avoid over-fitting. The initial grid of hyperparameters we use to perform the search is shown in Table~\ref{Tab:GridSearch_initial_grid}. We keep the remaining hyper-parameters in the RFR algorithm the same as the default values (these are listed in the lower part of Table~\ref{Tab:GridSearch_initial_grid}).

The parameters used for fine-tuning the model here are described as follows: 
\begin{itemize}
    \item "\texttt{n\_estimators}": the number of individual decision trees in the forest. 
    \item "\texttt{max\_features}": the maximum number of features considered when looking for best split at each node. The choices of \texttt{max\_features} includes ``\texttt{n}$_f$'' (all features or inputs available, which is \texttt{n}$_f$ = 6 in our case), ``\texttt{sqrt}'' (square root of the total features), and ``\texttt{log2}'' (log base 2 of the total features). The value of ``\texttt{sqrt}'' and ``\texttt{log2}'' after rounding down to the closest integer is the same in this work. Therefore the final hyperparameter choices are ``\texttt{n}$_f$'' and ``\texttt{sqrt}''. Using fewer features increases randomness, helping to de-correlate the trees. 
    
    % The number of features to consider when looking for the best split. This parameter can take on values of \texttt{n\_ features} or $\sqrt{\texttt{n\_ features}}$, where \texttt{n\_features} is a chosen parameter, limited by the number of observations. 
    \item "\texttt{max\_depth}": the maximum depth of the trees. If a value of ``None''  is chosen, then nodes are expanded until all leaves are pure (i.e. have the same class or value) or until all leaves contain less than "\texttt{min\_samples\_split}".
    \item "\texttt{min\_samples\_split}": the minimum number of samples required to split an internal node. 
    \item "\texttt{min\_samples\_leaf}": the minimum number of samples required to be at a leaf node. 
    \item "\texttt{min\_weight\_frac\_leaf}": the minimum weighted fraction of the total sum of weights required to be at a leaf node. 
    \item "\texttt{max\_leaf\_node}": the maximum number of leaf nodes.
\end{itemize}

\begin{table}
\centering
\begin{tabular}{ ccc} 
\hline
RFR  & Range Tested & Best Values \\
Hyperparameters & & \\
\hline
\texttt{n\_estimators}& [1,2,3,5,6,8,10,12,14,16,18,20, & 40\\ 
 & 25,35,40,50,75,100,125,150] & \\ 
\texttt{max\_features}& [\texttt{n}$_f$, ``\texttt{sqrt}''] & \texttt{n}$_f$\\ 
\texttt{max\_depth}& [``None'',1,2,4,6,8,10,15,] & 15\\
& 20,25,30,35,40,50] & \\
\texttt{min\_samples\_split}& [2, 5, 10] & 2\\
\texttt{min\_samples\_leaf}& [1, 2, 4] & 1\\
\hline
\texttt{min\_weight\_frac\_leaf}& & 0.0\\
\texttt{max\_leaf\_node} & & ``None''\\
\hline
\end{tabular}
\caption{The hyperparameters for the RFR. The first block of rows shows the tuned hyperparameters, the range of values or options tested (second column) and the best value adopted (third column). The parameters in the lower block were not varied. The hyperparameter names are explained in the text. }
\label{Tab:GridSearch_initial_grid}
\end{table}

In a formal grid search, there would be 5\,040 combinations of hyperparameters to test. Instead, a random sampling is made from the grid to speed up the hyperparameter search, using 200 parameter combinations.  
%Even though there are 5040 settings in total, the random grid search does not perform the search on all the settings. It selects the settings randomly to sample the wide range of values. This randomise is controlled by the number of iterations desired. The number of iterations is set to 200 for this work, meaning there are 200 combinations of settings randomly chosen for the grid search. \textbf{Can explain the above more succintly}
We also select the number of $k$-fold cross validations to be 7. This means that there are 1\,400 fitting iterations in total to be performed in the parameter search.
%By preforming the random grid search, we find that the best values for the RFR hyper-parameters are \textcolor{red}{insert the best values of the hyper-parameters here}

%With the best fine-tuned hyper-parameters, we train a machine learning model using the Random Forest Regression technique to learn the $k$-correction for each galaxy. We use the apparent magnitudes in $ugriz$ filters and the photometric redshifts. 

\subsubsection{The prediction error of the rest-frame absolute magnitude using the Random-Forest-Regression $k$-correction
}

\begin{figure}
\centering
    {\includegraphics[width=0.48\textwidth]{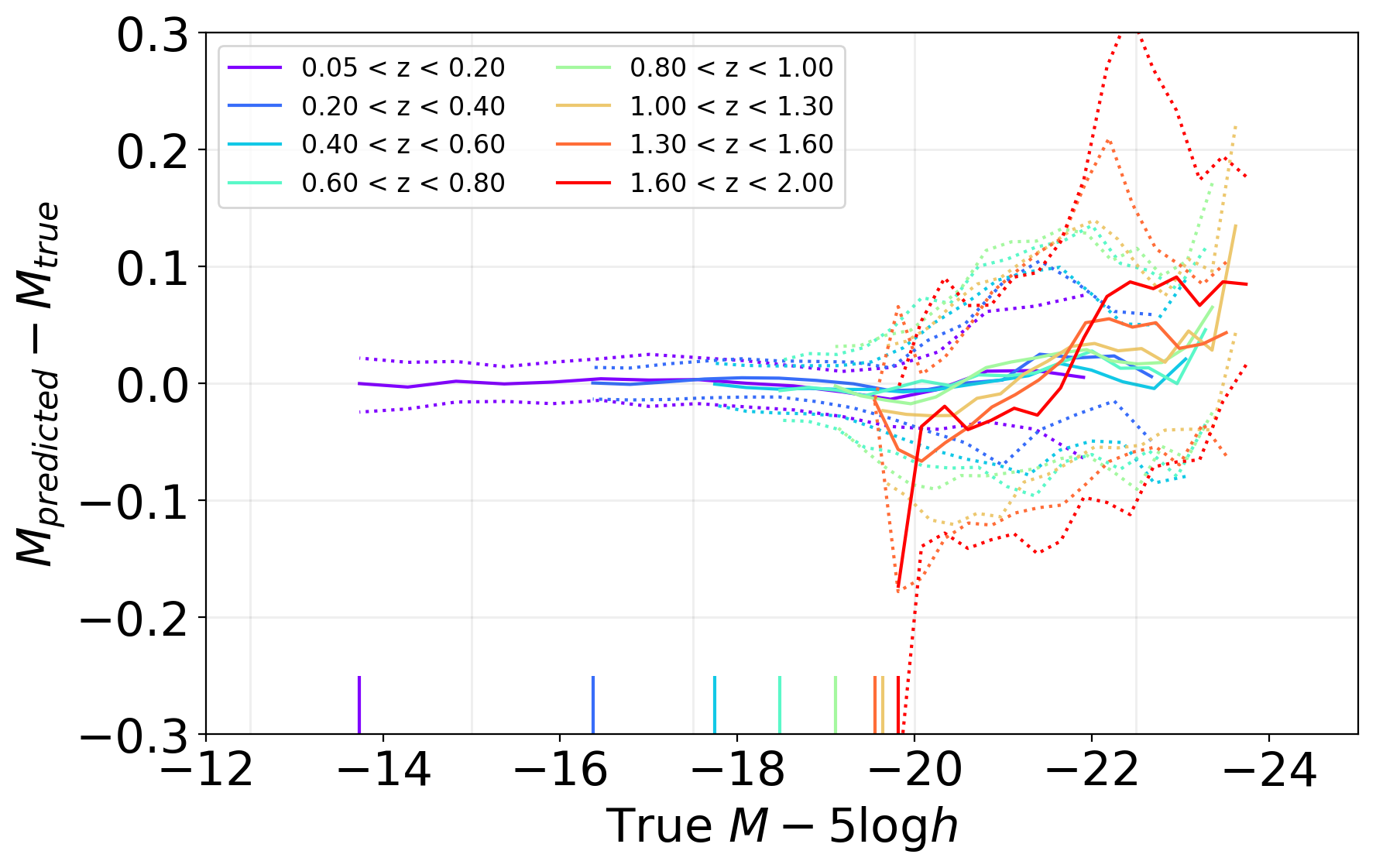}}%
\caption{The performance of the RFR machine learning estimate of the $k$-correction, expressed in terms of the difference between the predicted and true absolute magnitude for each galaxy. The true magnitude is predicted using the exact $k$-correction from \texttt{GALFORM}. The solid lines show the median difference between the true and predicted magnitudes. The dotted lines show the 16-84 th percentile interval, a centralised version of the $1$-$\sigma$ scatter which is not affected by outliers. The short vertical lines at the bottom of the panel show the absolute magnitude limits per redshift bin. Different colours indicate different redshift ranges, as shown by the legend.}%
    \label{fig:MagnitudeError_ML}%
\end{figure}

The RFR machine learning prediction of the $k$-correction is compared to the exact answer from \texttt{GALFORM} in Fig.~\ref{fig:MagnitudeError_ML}. The error is presented as the difference between the predicted and true absolute magnitudes.\footnote{Throughout absolute magnitudes are referred to the Hubble-parameter independent magnitudes as $M - 5\log h$ to facilitate comparison with literature derived using different $h$ values: this means that distances are expressed in units of $h^{-1}\textrm{Mpc}$.} A magnitude difference of zero indicates that the machine learning method reproduces the \texttt{GALFORM} answer exactly. The results are shown for different redshift slices, as indicated by the legend. For $z \le  1$, there is no bias in the predicted absolute magnitude, just a small scatter which reaches $\approx 0.05$ magnitude for the brightest galaxies. At $z>1$, there is a bias in the recovered magnitude, with the median differing from zero by up to $\pm \,0.2$ magnitude. The scatter is also larger, $\sim 0.1$ magnitude. This sign of the bias means that the brightest galaxies are assigned an estimated absolute magnitude that is too faint, whilst the faintest galaxies in the highest redshift slices are brighter than they should be.   

\section{Estimating the luminosity function from the mock catalogue}

We estimate the galaxy luminosity functions (LFs) using the $V_{\text{max}}$ methodology \citep{Schmidt1968}. In a flux-limited sample, the number of sources observed varies strongly with luminosity \citep{Driver1996}. At the faint end, few galaxies are seen because they are only brighter than the apparent magnitude of the selection limit at low redshifts. So, even though we expect to find many faint galaxies per unit volume, they are only visible over a small volume compared to brighter galaxies. The number of bright galaxies observed is also small, even though these are visible to much greater redshifts than faint galaxies. This is due to the intrinsic luminosity distribution of galaxies, with the space density of bright galaxies falling exponentially within increasing luminosity brightwards of a characteristic luminosity, $L_*$. The observed number of galaxies therefore peaks around $L_*$. The  $V_{\text{max}}$ approach corrects the observed number of objects to provide an estimate of the true, underlying number density of galaxies, by accounting for the volumes that galaxies of different luminosities can be observed over. 

The value of $V_{\text{max}}$ is given by the maximum redshift out to which the galaxy could be seen. Starting at the redshift of observation, $z$, we can imagine increasing the redshift gradually, adjusting the luminosity distance and $k$-correction accordingly until we reach the maximum redshift, $z_{\rm max}$, at which the galaxy could still be observed. At this redshift, the apparent magnitude of the galaxy is equal to the limit that defines the catalogue, $m_{\rm limit}$: 
\begin{equation}
m(z_{\rm max}) = m_{\rm limit} = M + 5 \log(d_{\rm L}(z_{\rm max})/{\rm Mpc}) + k(z_{\rm max}) + 25,
\label{eq:DM}
\end{equation}
where $d_{\rm L}$ is the luminosity distance. 

\begin{figure}
\centering
    {\includegraphics[width=0.45\textwidth]{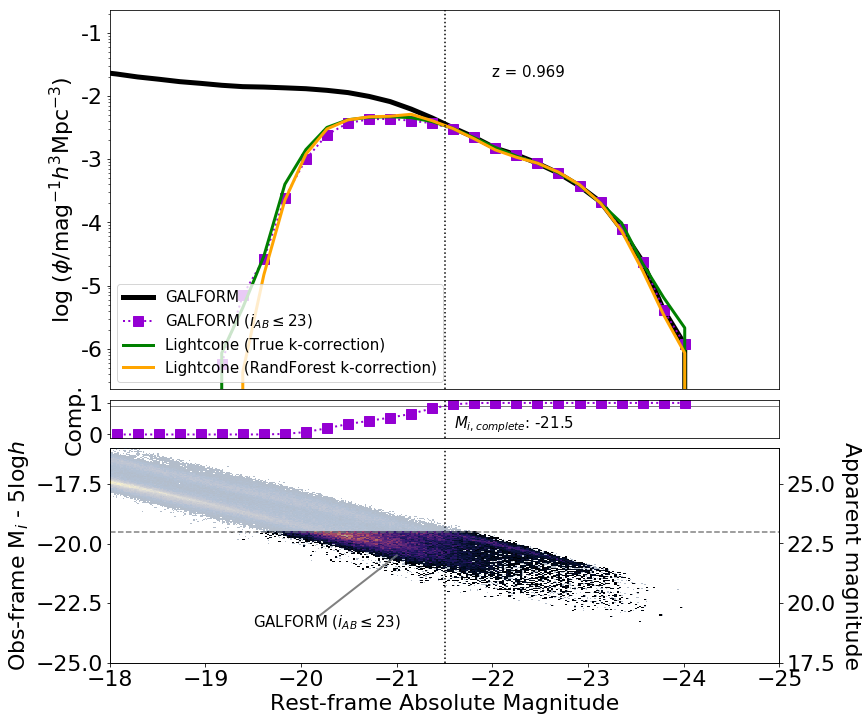}}%
 \caption{The impact of selection effects on the estimated luminosity function, shown at $z \sim 1$ for illustration. \textit{(top):} The black curve shows the \texttt{GALFORM} LF without \textit{any} selection effects. The purple points connected using a dotted line show the LF after applying the observed $i$-band limit of $i_{\rm AB}=23$, using a weighted average of the LFs for the snapshots that fall within the redshift range of the shell around $z \sim 1$. The weighting is given by ${\rm d}N/{\rm d}z$. The green curve shows the LF estimated from the lightcone mock, assuming the exact $k$-correction predicted by \texttt{GALFORM}. The orange line shows the LF recovered using the $k$-correction obtained using the random forest and the $ugriz$ photometry. \textit{(middle):} The purple points show the ratio between the \texttt{GALFORM} LF with selection effects and the true \texttt{GALFORM} LF. The LF is considered complete down to $M_i = -21.5$ at which point the recovered LF falls to 90 per cent of the true LF, as indicated by the vertical dotted line across three panels. \textit{(bottom):} The density plot shows the observed \textit{vs} rest $i$-band absolute magnitudes of the model galaxies; points below the horizontal dotted line pass the sample selection. }%
    \label{fig:GALFORM_LF_selection}
\end{figure}

PAUS galaxies are visible over a wide range in redshift (see Fig.~\ref{fig:redshift_distribution} and see also \citealt{Manzoni2024}). Given the relatively large solid angle probed, compared to other deep surveys, we can divide the volume covered into a series of thin redshift shells to isolate evolution in the luminosity function. Due to the use of thin shells in redshift, most galaxies are visible over the full redshift interval of the shell, $z_{1} < z <z_{2}$, so $z_{\rm max}=z_{2}$. This means that for most galaxies in each shell, $V_{\text{max}} = V_{\rm shell}$. It is only for the faintest galaxies that $z_{\rm max} < z_2$, as these galaxies are only visible over part of the shell. Note that since we know the value of the $k$-correction at the redshift of observation, but we do not know the functional dependence on redshift, we make the approximation that the $k$-correction is constant when perturbing the redshift to find $z_{\rm max}$. Effectively, this means $k(z_{\rm max}) = k(z)$, where $z$ is the redshift at which the galaxy crosses the observer's past lightcone. Again, this assumption only affects a small number of faint galaxies in each redshift slice. 

We first focus on the estimated luminosity function in a thin redshift slice at $z\sim1$ to illustrate some features of our analysis. Fig.~\ref{fig:GALFORM_LF_selection} shows various estimates of the luminosity function of the light-cone mock compared to the original prediction from \texttt{GALFORM}. The target or true luminosity function in this example is the  \texttt{GALFORM} prediction from the simulation box, which is simply a histogram of all the galaxies in the simulation volume, binned in luminosity, without any consideration of whether or not the galaxy is bright enough to meet a selection limit in the observed $i$-band. Hence, for this prediction, the number of galaxies keeps rising as the luminosity bin gets fainter (eventually, if we consider a faint enough bin, the \texttt{GALFORM} luminosity function will turn over due to mass resolution effects in the N-body simulation). The curve labelled \texttt{GALFORM} in Fig.~\ref{fig:GALFORM_LF_selection} is actually a weighted combination of the snapshot predictions that fall within the redshift shell, with the weight applied to each LF being the redshift distribution of galaxies at each redshift. 

After taking into account the magnitude limit, $i_{\rm AB}=23$, the LF estimated from the lightcone mock is shown by the purple symbols with dotted line in the upper panel of  Fig.~\ref{fig:GALFORM_LF_selection}. Rather than showing a sharp cut in the LF in the rest-frame $i$-band, there is a gradual reduction in the amplitude of the luminosity function  as we move to fainter magnitudes.  The reason for this is that the sample selection is in the observed $i$-band, whereas we plot the estimated LF in the rest-frame $i$-band. This point is illustrated further in the lower panel of Fig.~\ref{fig:GALFORM_LF_selection} which shows the observed-frame $i$-band absolute magnitude (i.e. the absolute magnitude in Eq.~\ref{eq:DM}, but without applying the $k$-correction) plotted against the rest-frame $i$-band, down to apparent magnitudes much fainter than $i_{\rm AB}=23$. The overall distribution of galaxies is plotted as a light 2D density histogram, with the (intrinsically) red and blue populations visible as the ridges of the density histogram. The galaxies which meet the observed $i_{\rm AB}=23$ selection are plotted a darker 2D histogram, labelled as ``GALFORM $(i_{\rm AB} \leq 23)$''. Galaxies with red observed $g-r$ colours make it into the sample over a wider range of rest-frame $i$-band magnitude than blue galaxies. 

The underlying luminosity function estimated over the magnitude range of the turn over is incomplete and is driven by the survey selection and the colour distribution of galaxies. Hence, there is still useful information over this magnitude range which can be used to constrain galaxy formation models, if the same selection effects can be applied to the model galaxies, as is the case with our mock catalogue. The faintest rest-frame magnitude bin of the luminosity function to be considered complete is indicated with the vertical dashed line in Fig.~\ref{fig:GALFORM_LF_selection}. The completeness limit is defined as the magnitude at which the $i_{AB}=23$ flux-limited \texttt{GALFORM} luminosity function remains above 90 per cent of the underlying true \texttt{GALFORM} luminosity function, as shown in the middle panel of Fig.~\ref{fig:GALFORM_LF_selection}.

In the redshift bins closer to $z \sim 0$, the observed and rest frame $i$-bands are closer together in redshift and the turnover at faint magnitudes in the recovered LF is narrower than it is at $z \sim 1$.

\section{Estimating the luminosity 
function from observations }

In this section we present our main results. In \S~4.1 we show the measurements of the $i$-band LF from PAUS. These results are compared to previous measurements from the literature in \S~4.2. In \S~4.3 we show the LF measured for red and blue galaxies. Finally, in \S~4.4 we discuss the completeness of the samples and the impact of making cuts on the quality of the photometric redshifts. 

\subsection{$i-$ band luminosity function - all galaxies}

\begin{figure*}
\centering
    {\includegraphics[width=0.75\textwidth]{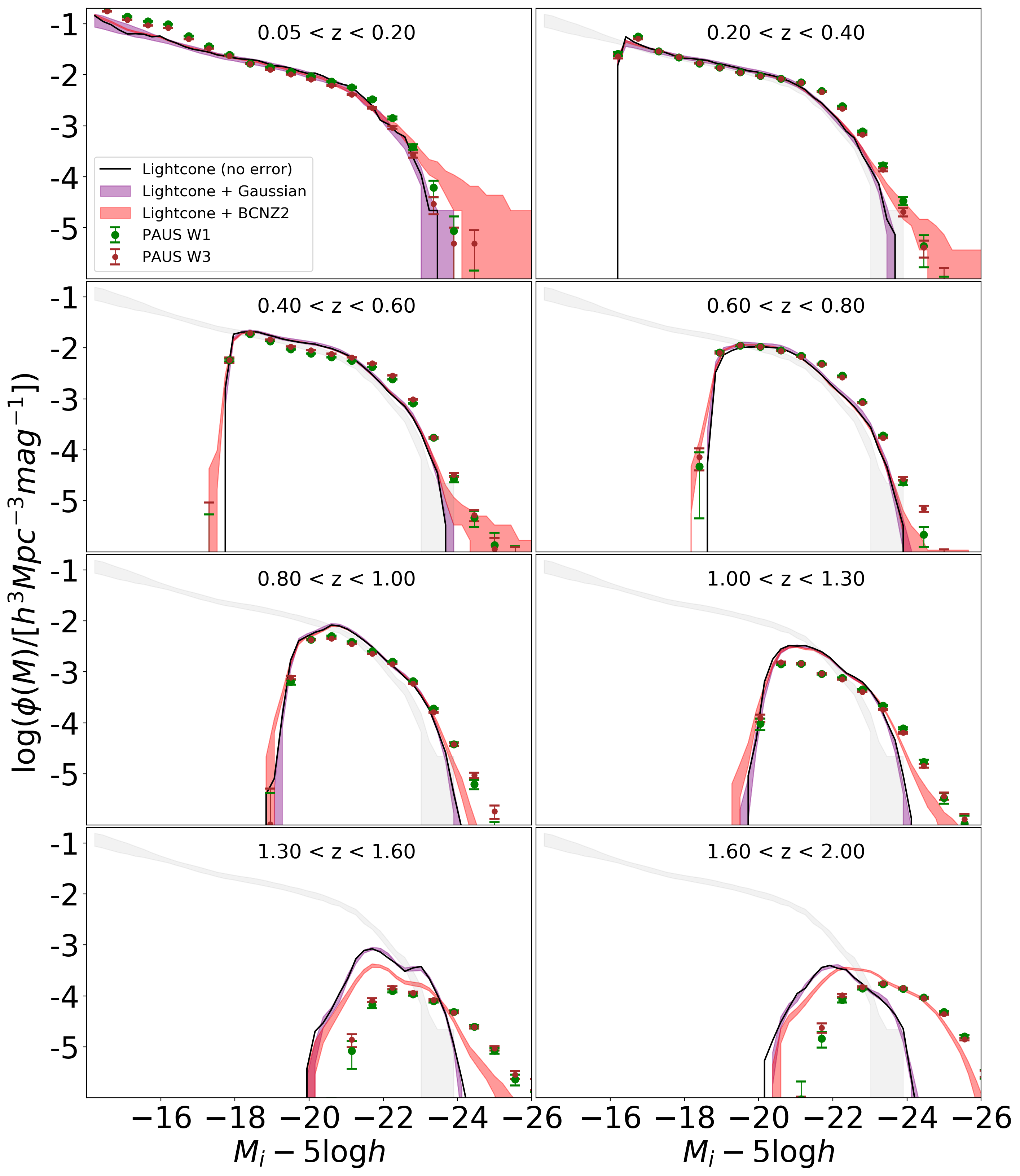}}%
\caption{The rest-frame $i-$band galaxy luminosity function for redshift bins ranging from $z = 0.05$ to $z = 2.00$. Each panel corresponds to a specific redshift range, as indicated at the top of each subplot (see Fig.~\ref{fig:redshift_distribution} for the corresponding redshift distribution). The measurements are shown for two PAUS fields, W1 (green points) and W3 (red points). These are compared with predictions from two variants of \texttt{GALFORM} mock catalogue, Lightcone + Gaussian photometric redshift method (purple shaded region) and Lightcone + \texttt{BCNZ}2 photometric redshift method (pink shaded region), which are explained in the text in \S~4.1. The lightcone luminosity function with no errors is plotted as a black solid line in each panel for reference.}%
    \label{fig:LC_PAUS_W1_W3}%
\end{figure*}  
% The grey shading shows the Lightcone + Gaussian LF for redshift $0.05 < z < 0.20$ for reference.

We compare the galaxy luminosity function estimated from the \GALFORM lightcone mock catalogue of \cite{Manzoni2024} with that obtained from observational data in the PAUS W1 and W3 fields (\citealt{Navarro2024}), using the $V_{\rm max}$ methodology. The \GALFORM mock catalogue serves as a baseline to test our luminosity function estimates revealing the impact of uncertainties and systematic effects introduced by the survey selection and methodology. These uncertainties, arising from the photometric errors, large-scale structure sampling variance, and photometric redshift errors, are incorporated into the mock catalogue following the approach described in \S2.3, and can be switched on and off to show how they affect the recovered luminosity function. 

In this analysis, we consider two treatments of photometric redshift errors. In one case, we apply the error distribution obtained from applying the \texttt{BCNZ}2 code to a random sample of the mock galaxies. This approach incorporates outliers in the photometric redshifts. In the other case, we mimic the central distribution of photometric redshift errors using Gaussian-like perturbations based on the $\sigma_{68}(\Delta z)$ values reported by \cite{Navarro2024}. The latter method ignores outliers that lie outside the tails of the Gaussian. This dual approach allows us to explore the impact of low-quality photometric redshift measurements (or outliers) on the recovered luminosity function. Later on we will investigate the effect on the estimated luminosity function of making a cut on the quality factor of the photometric redshifts. 

The uncertainties in the estimated luminosity function are calculated as follows: (1) large-scale structure or sampling variance errors are estimated using the jackknife method with $N_{\rm regions} = 64$ (following \citealt{Norberg+2009}), (2) the contribution of photometry errors and photometric redshift errors to the error on the recovered LF is derived via a Monte Carlo technique. We generate 500 realisations of the mock, including both sources of error, and estimate the LF in each case. The error is the 1-$\sigma$ range of the LF estimates, which is shown as error bars in Fig.~\ref{fig:LC_PAUS_W1_W3}. In particular, the combined photometric and photometric redshift errors are approximately 1 dex larger than the large-scale structure error across all redshifts. 

\begin{figure*}
\centering
    {\includegraphics[width=0.99\textwidth]{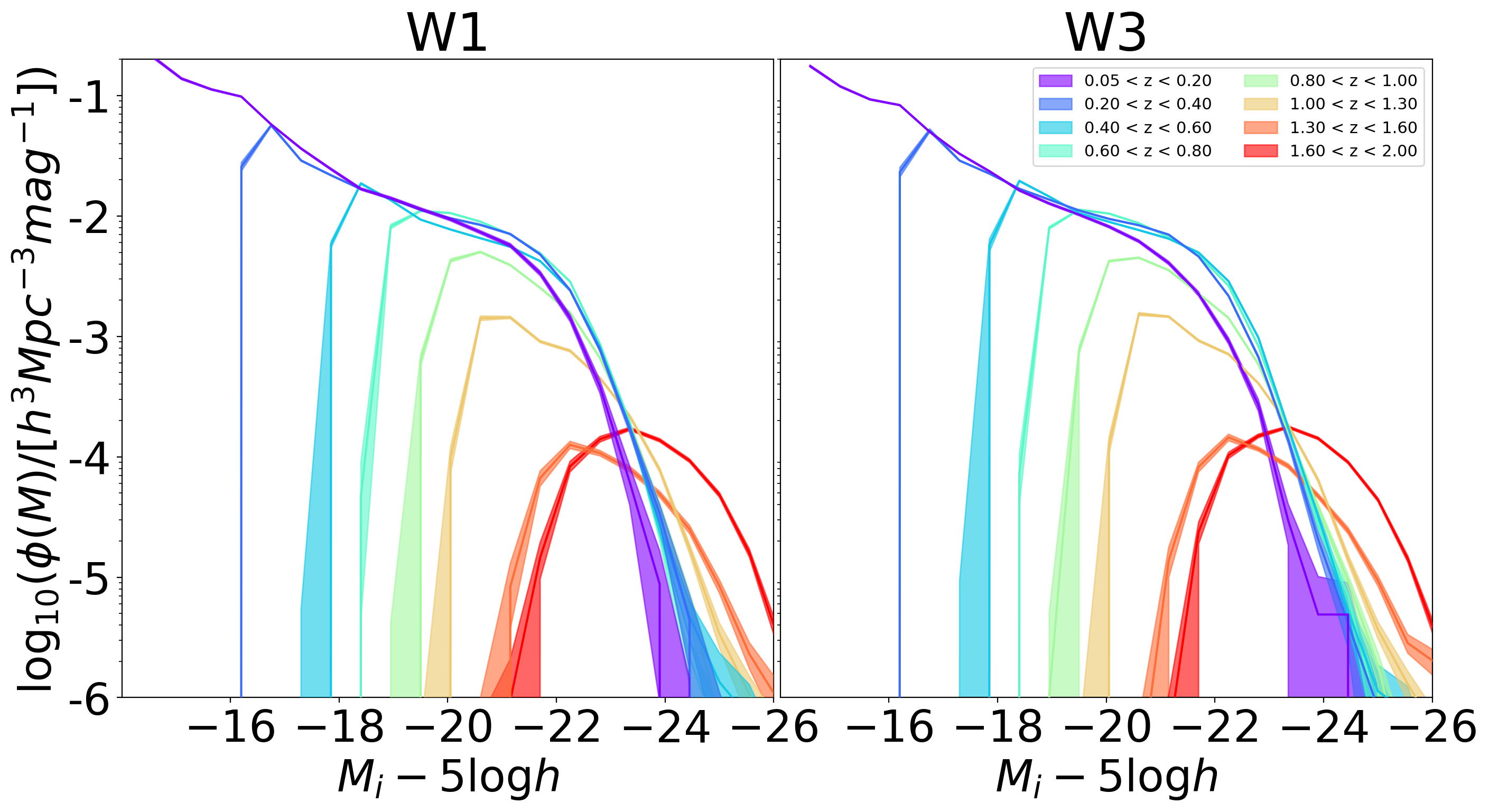}}%
\caption{The galaxy luminosity function estimated from the PAUS W1 (left) and W3 (right) fields, focusing on the evolution with redshift. The LFs from the different redshift slices are plotted on a single panel for each field, as indicated by the legend in the right panel. Red colours correspond to low redshift and blue colours to high redshift.}%
    \label{fig:LC_PAUS_W1_W3_evol}%
\end{figure*}

\begin{figure*}
\centering
    {\includegraphics[width=\textwidth]{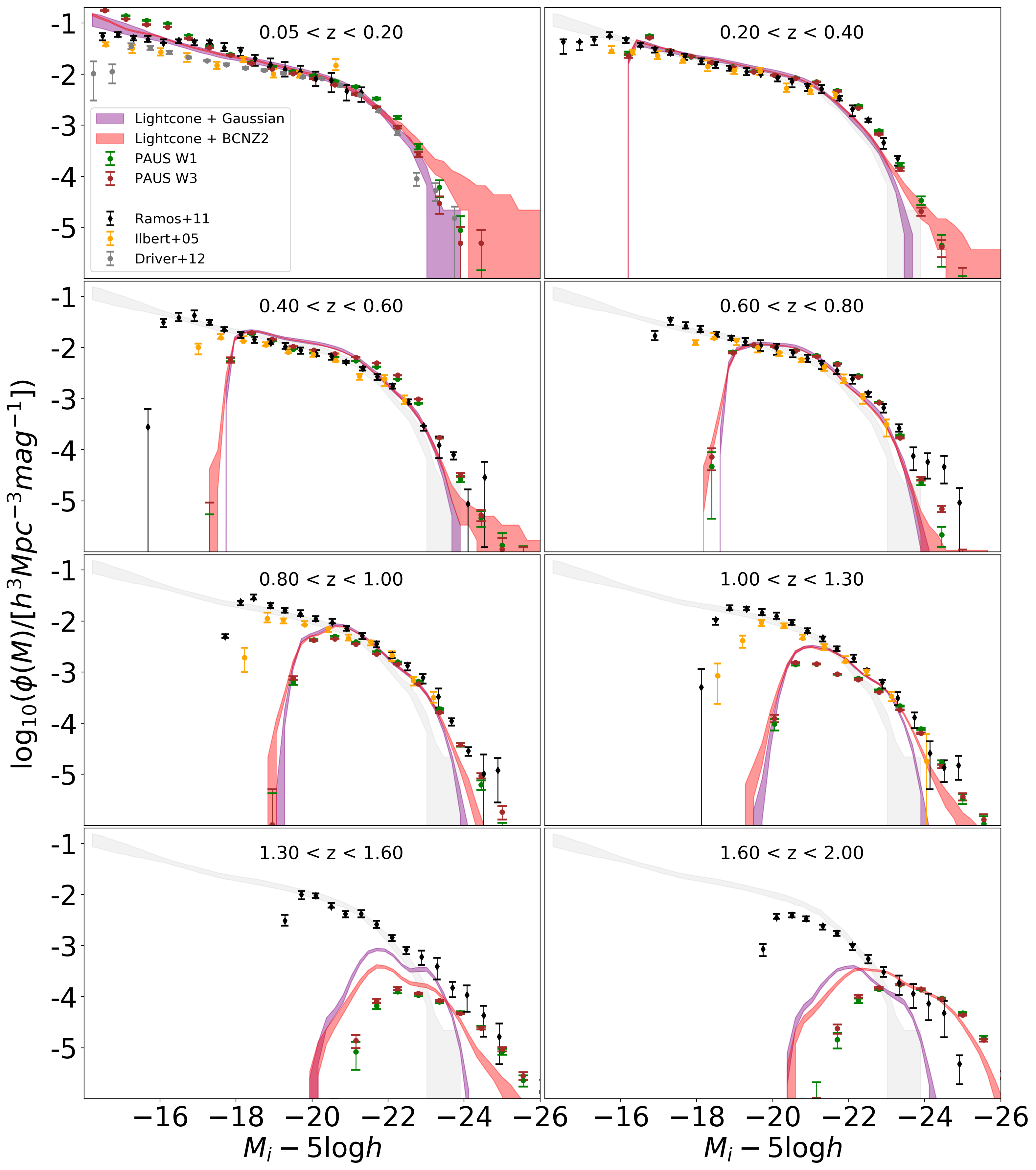}}%
\caption{The galaxy luminosity function estimated in the rest frame $i$-band in 8 redshift bins, compared with the predictions from the lightcone and previous estimates, as indicated by the key in the top left panel. The mock catalogue and W1 and W3 fields are limited to $i=23$. The selection for the other surveys is given in the text. The lightcone results for the lowest redshift bin are reproduced in the other panels as the grey shading for reference.}
    \label{fig:LC_PAUS_W1_W3_ramos}%
\end{figure*} 

In Fig.~\ref{fig:LC_PAUS_W1_W3} we present the rest-frame CFHTLS $i$-band luminosity function\footnote{The LF in other bands is shown in Appendix A} in redshift bins over the redshift range $0.05 < z < 2.0$ (see the redshift distributions in Fig.~\ref{fig:redshift_distribution} for reference to help interpret these results). In each redshift bin, the purple shaded region corresponds to the mock luminosity function when Gaussian-like photometric redshift errors are applied, whilst the pink shaded region shows the effect of the more realistic, \texttt{BCNZ}2-like photometric redshift uncertainties, which include photometric redshift outliers. Observational data are plotted using green filled diamonds for the PAUS W1 field and red solid circles for PAUS W3. 

% \textcolor{red}{the black solid line represents the "true" luminosity function from the lightcone mock catalogue - i.e. the LF including the $i$-band selection but without any errors in the photometry or photometric redshifts}. 

Fig.~\ref{fig:LC_PAUS_W1_W3} illustrates the effect of photometric redshift measurement errors on the estimated luminosity function. With Gaussian-like errors and no formal outliers (i.e. no redshift errors that lie beyond the wings of a Gaussian distribution), the errors in redshift lead to errors in luminosity but these are small compared to the size of the luminosity bins used to plot the LF. Hence, the overall LF fluctuates without a significant change in shape (purple shading). In contrast, when the more realistic photometric redshift errors are applied, including outliers, the LF shape is markedly different, especially at the bright end, where the break becomes less pronounced. 
This occurs because a large error in redshift can lead to a large change in luminosity which places a galaxy in a different luminosity bin; this can lead to an appreciable change in the shape of the LF at the bright end, where the variation of number density with increasing luminosity is rapid. At the faint end, the number density of galaxies varies more slowly with luminosity, so errors in luminosity have less impact on the shape of the LF.

Up to $z \sim 1$ (equivalent to a lookback time of $\sim~8.5$~Gyr, more than 60 per cent of the age of the Universe), the luminosity function predicted from the \GALFORM mock catalogue agrees reasonably well with the observational estimates from PAUS W1 and W3. At higher redshifts, the introduction of realistic \texttt{BCNZ}2-like photometric redshift errors improves the match at the bright end. In particular, although there is an offset, the shape of the LF at the bright end estimated from PAUS is similar to that recovered from the mock, when the photometric redshift outliers are included. However, discrepancies at the faint end become apparent from $z \sim 1$ onward, growing larger at higher redshifts.

We do not attempt to fit a parametric form to the measured luminosity function, as our goal is to directly compare the observed LF with the mock catalogue that shares the same selection effects. For instance, fitting a single Schechter function to our results would require discarding magnitude bins near the turnover at the faint end, which contain valuable information for testing galaxy formation models. 

We now focus on the evolution in the LF estimated from the observations. Fig.~\ref{fig:LC_PAUS_W1_W3_evol} shows the evolution of the LF by plotting the estimates in different redshift bins in the same panel. Note that whereas we have shown observational estimates of the LF using symbols in other plots, in this figure we have switched to using curves. The thickness of the curve shows the estimated error on the LF estimate. The results are shown separately for the W1 (left panel) and W3 (right panel) fields. The main evolutionary effect we see is the impact of the apparent magnitude selection in the observed $i$-band. With increasing redshift, the turnover at faint magnitudes shifts to brighter magnitudes and becomes less sharp. The latter effect is due to the increasing difference between the observed and rest frame $i$-band with increasing redshift. Otherwise, the faint end of the $i$-band LF changes little. There is a modest brightening of the LF around the break around $z \sim 0.5$. The biggest change at the bright end is at higher redshift, with a reduction in the sharpness of the break at the bright end, mainly due to photometric redshift outliers.

\subsection{Comparison with previous estimates of the LF}

Here we compare our estimate of the $i$-band LF from PAUS with results from other surveys. These surveys includes studies from \cite{Ilbert2005}, \cite{Ramos2011}, and \cite{Driver2012}.
% The redshift slices in Fig.~\ref{fig:LC_PAUS_W1_W3_ramos} show broad intervals in redshift, which contain the redshifts used in the other surveys. However, the width of the redshift slices in the other surveys may not match that used here. 

\cite{Ilbert2005} measured spectroscopic redshifts for 11\,000 galaxies to $i_{\rm AB}=24$ in the VIMOS VLT Deep Survey. In view of this difference in the depth of the survey compared with PAUS, the turnover in the Ilbert et~al. LF estimates should appear about 1 magnitude fainter than in the estimates from PAUS. The Ilbert et~al. estimate is from a relatively small solid angle and so does not extend to as bright a magnitude as the PAUS LF estimates. Also, Ilbert et~al. impose a bright magnitude cut of $i_{\rm AB}=17.5$. \cite{Ramos2011} measured the LF from the CFHTLS deep fields, covering in total just under 3 square degrees to depths close to $i_{\rm AB}=26$. These authors use photometric redshifts, derived from the broad band photometry of CFHTLS. Hence, both the scatter and outlier fractions for their photometric redshifts are expected to be larger than those in PAUS. The Ramos et~al. estimates extend to the faintest rest-frame $i$-band absolute magnitude as expected from their deeper apparent magnitude limit. At the bright end, the estimates from Ramos et~al. are affected by sample variance and the errors in the photometric redshifts. Finally, \cite{Driver2012}  used the Galaxy And Mass Assembly survey to measure the LF in many bands. Driver et al. use spectroscopic redshifts. Their sample is selected in the $r$ band to depths of $r_\textrm{Kron}=19.4$, compared to $i_\textrm{AB} = 23$ in our work. We show a low redshift estimate from Driver et~al. (top left panel). 

\begin{figure*}
\centering
    {\includegraphics[width=0.7\textwidth]{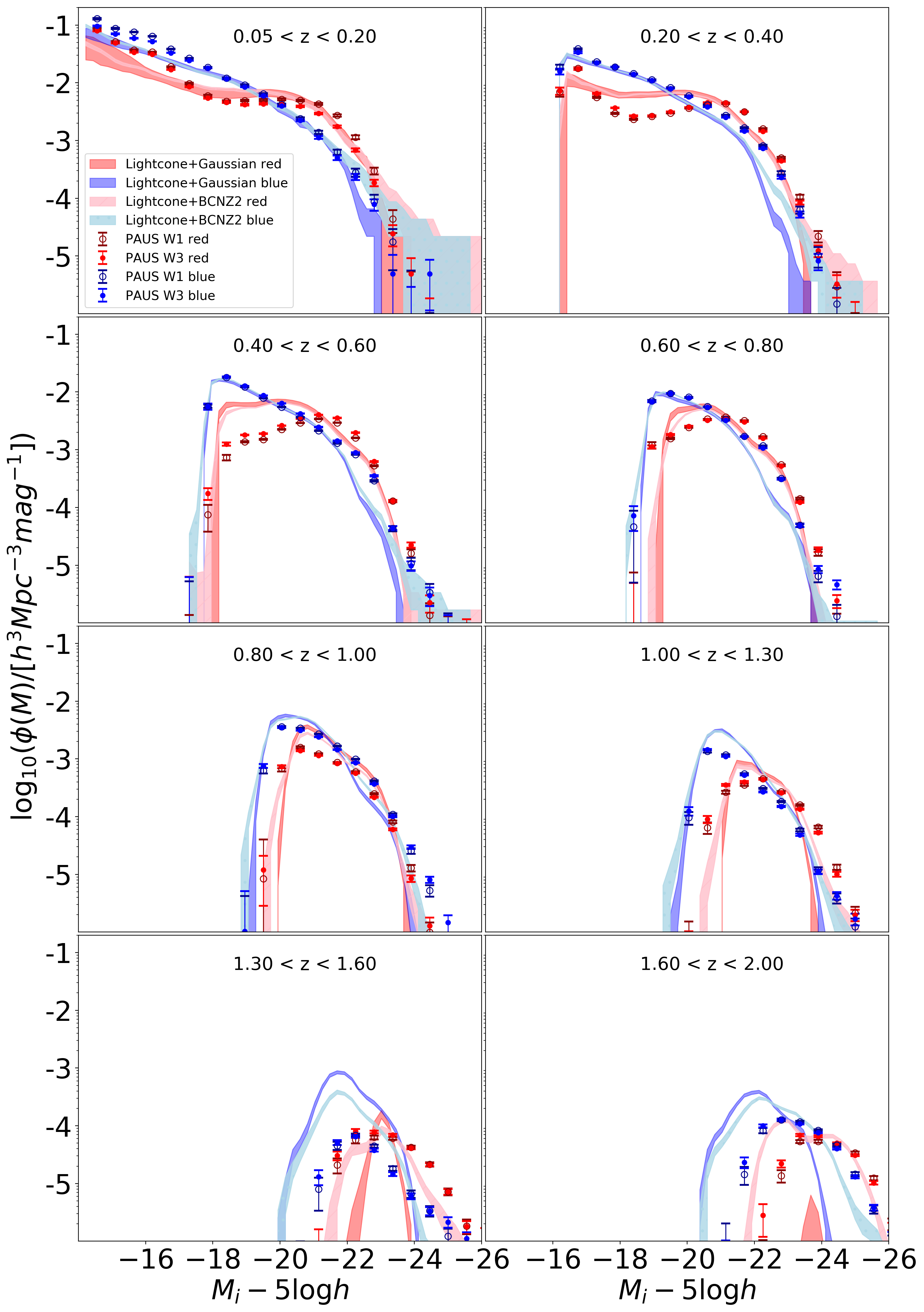}}%
\caption{The $i-$band galaxy luminosity function for red and blue galaxies (as defined in the observed-frame) for multiple redshift bins ranging from $z = 0.05$ to $z = 2$. Each panel corresponds to a specific redshift range, as indicated at the top of each panel (see Fig.~\ref{fig:redshift_distribution} for the corresponding redshift distribution). The measurements are shown for the two PAUS fields, W1 (open points) and W3 (filled points). These are compared with predictions from two variants of GALFORM mock catalogue, Lightcone + Gaussian photometric redshift method (darker shaded region) and Lightcone + \texttt{BCNZ}2 photometric redshift method (lighter shaded region); in both cases the points and the shading are based on the colour of the sample.}%
    \label{fig:LC_PAUS_W1_W3_red_blue}%
\end{figure*}

\begin{figure*}
\centering
    {\includegraphics[width=0.9\textwidth]{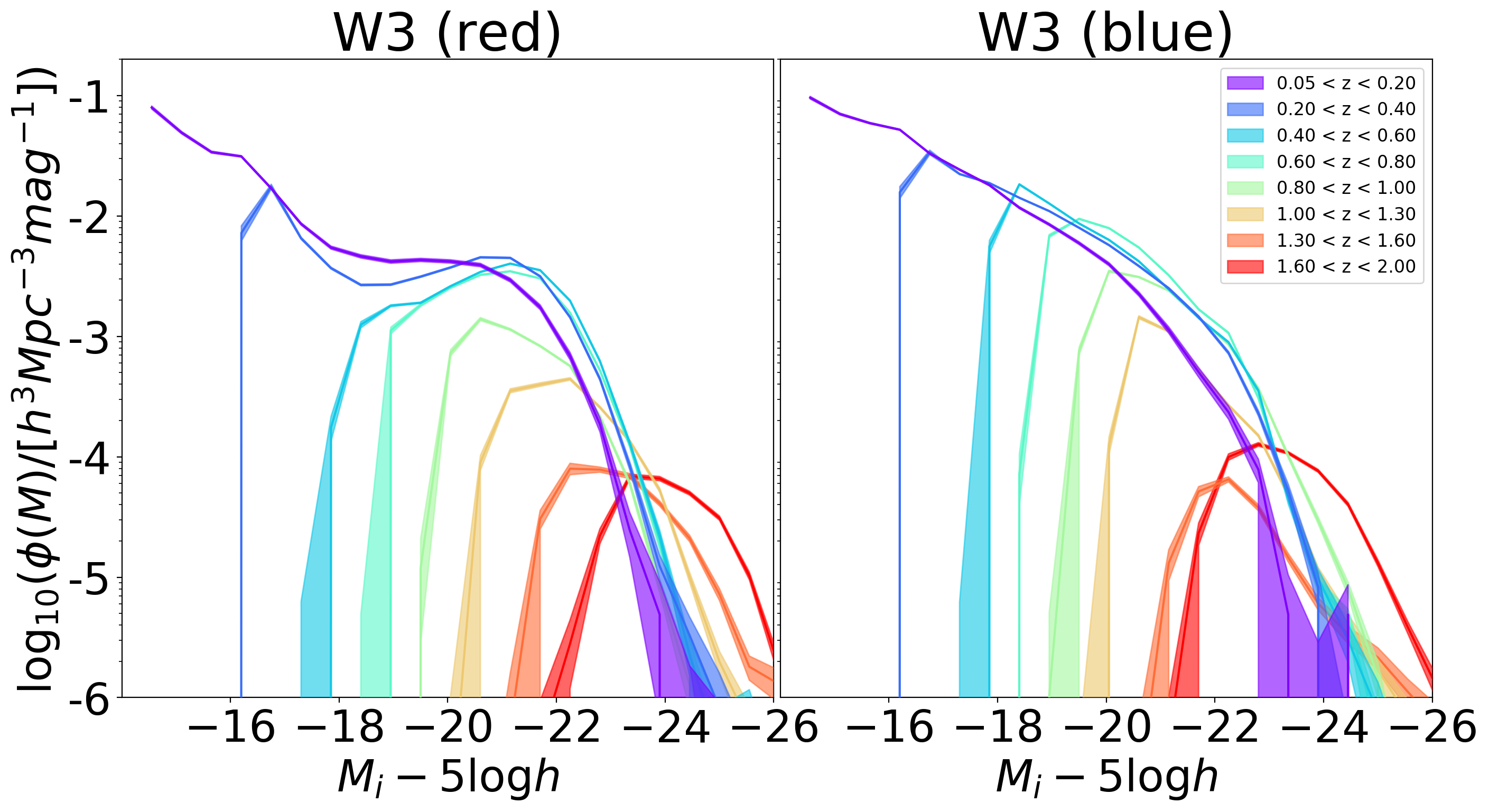}}%
\caption{The $i-$band galaxy luminosity function for observed-frame red (left panel) and blue (right panel) galaxies for multiple redshift bins ranging from $z = 0.05$ to $z = 2$, as indicated by the legend. Here we focus on the larger of the PAUS fields, W3. The observational estimates are shown by curves rather than symbols, with the width of the curve indicating the $1-$$\sigma$ error on the measurement.}%
    \label{fig:LC_PAUS_red_blue_evol_W3}%
\end{figure*}

The differences in the LFs from the literature and our new estimates can be readily understood at the faint end as being due to the different $i$-band cuts applied, as outlined above. The differences at the bright end mainly reflect the smaller fields used in Ramos et~al. and Ilbert et~al., which limit how bright their estimate can reach. At intermediate magnitudes there is reasonable agreement between the different estimates.

\subsection{The red and blue galaxy luminosity functions}

We now delve more deeply into what is driving the evolution in the overall LF, by following \cite{Lilly1995} and looking at the LF for red and blue galaxies. Whereas Lilly et al used the rest-frame colour to label galaxies as red or blue, here we follow \cite{Manzoni2024} and use the observed $g-r$ colour. In practice this means that the dividing line between red and blue galaxies is a function of redshift, rather than a constant as would be the case for a rest-frame colour. 
The LF of red and blue galaxies is shown in Fig.~\ref{fig:LC_PAUS_W1_W3_red_blue}. Here we also plot the red and blue galaxies from the PAUS mock. For the mock, two versions of the photometric redshift errors are considered - the one labelled `Gaussian' which neglects outliers and the other labelled `BCNZ2' which does attempt to include outliers. Note that the photometric redshift error distribution is sampled for an $i$-band selected sampled, and does not take into account galaxy colour. The version of the mock errors labelled `BCNZ2' is  the more realistic, and gives a boosted number of bright galaxies. 

Qualitatively, the observed and predicted LFs of red and blue galaxies agree. The turnover at faint magnitudes is arguably less well reproduced for red galaxies than for blue galaxies; this difference becomes more apparent with increasing redshift. 

We take another view of the evolution of the red and blue LF in Fig.~\ref{fig:LC_PAUS_red_blue_evol_W3}. Here we just show the observed LF estimated in the bigger of the PAUS fields, W3. No model LFs are plotted in this figure to allow us to focus on how the observational estimate of the LFs evolve. There is more evolution in the shape of the faint-end of the red LF at low redshift, than in the blue LF, which retains a power-law shape. At intermediate and high redshifts, we see a shift in the LF from $L_*$ and brighter. This shift is more pronounced for blue galaxies than red ones. At the very highest redshifts, the break in the LF becomes less severe due to photometric redshift outliers. 
We note that a morphological separation could provide a more physically motivated classification; however, such measurements are not currently available for the PAUS data. The colour-based criteria used in this work is applied consistently to both the observation data and the mock catalogue.

\begin{figure*}
\centering
    {\includegraphics[width=0.9\textwidth]{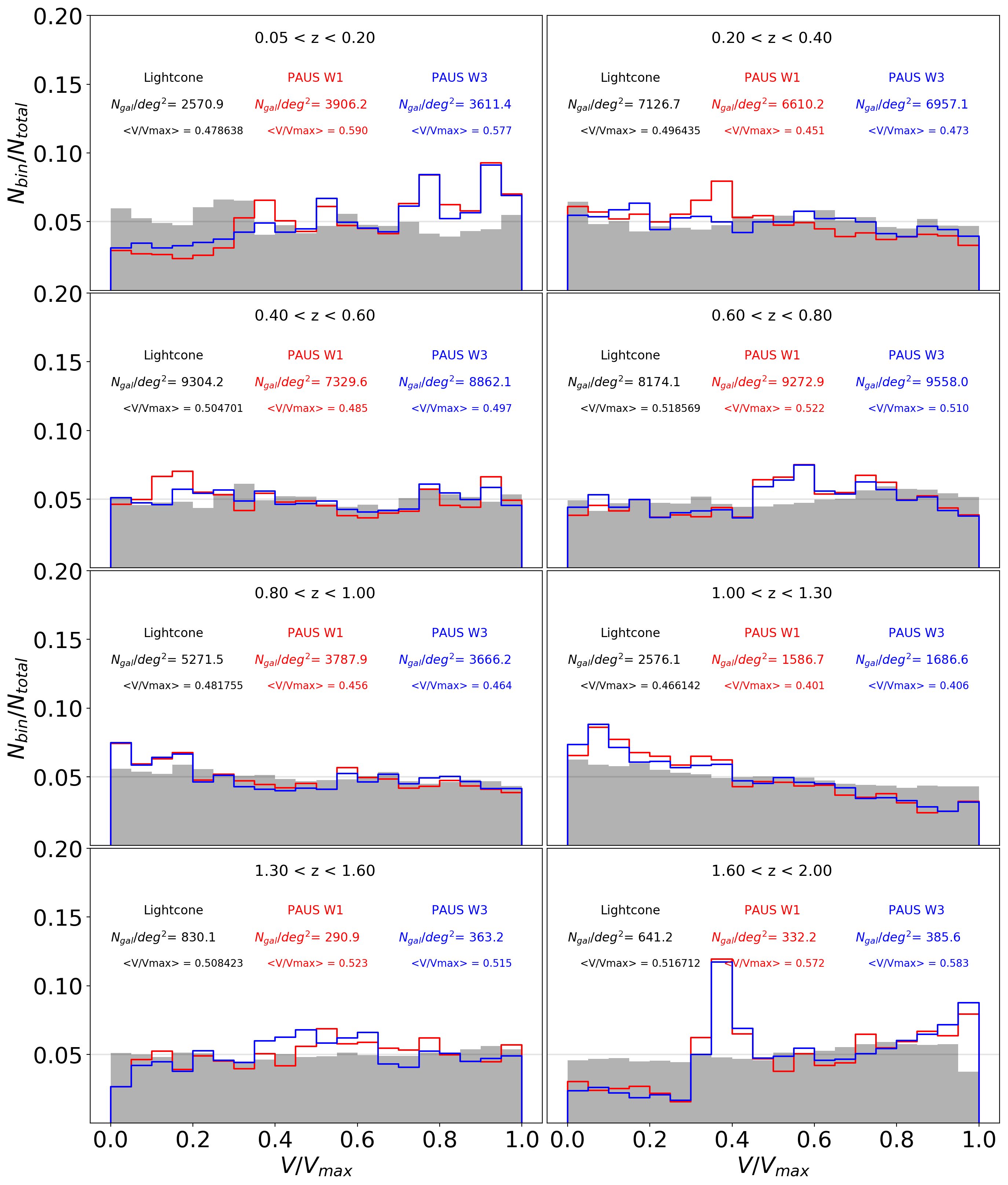}}%
\caption{The distribution of $V/V_{\textrm{max}}$ for different redshift bins, as labelled in each panel. Distributions are shown for the mock lightcone catalogue (grey histogram), without any errors in the photometry or photometric redshifts of the model galaxies, the PAUS W1 (red histogram) and W3 (blue histogram) fields. The labels give the number of galaxies in each case and the mean value of $\langle V / V_{\textrm{max}}\rangle$.}%
    \label{fig:VVmax_all}%
\end{figure*}

\begin{figure}
\centering
    {\includegraphics[width=0.49\textwidth]{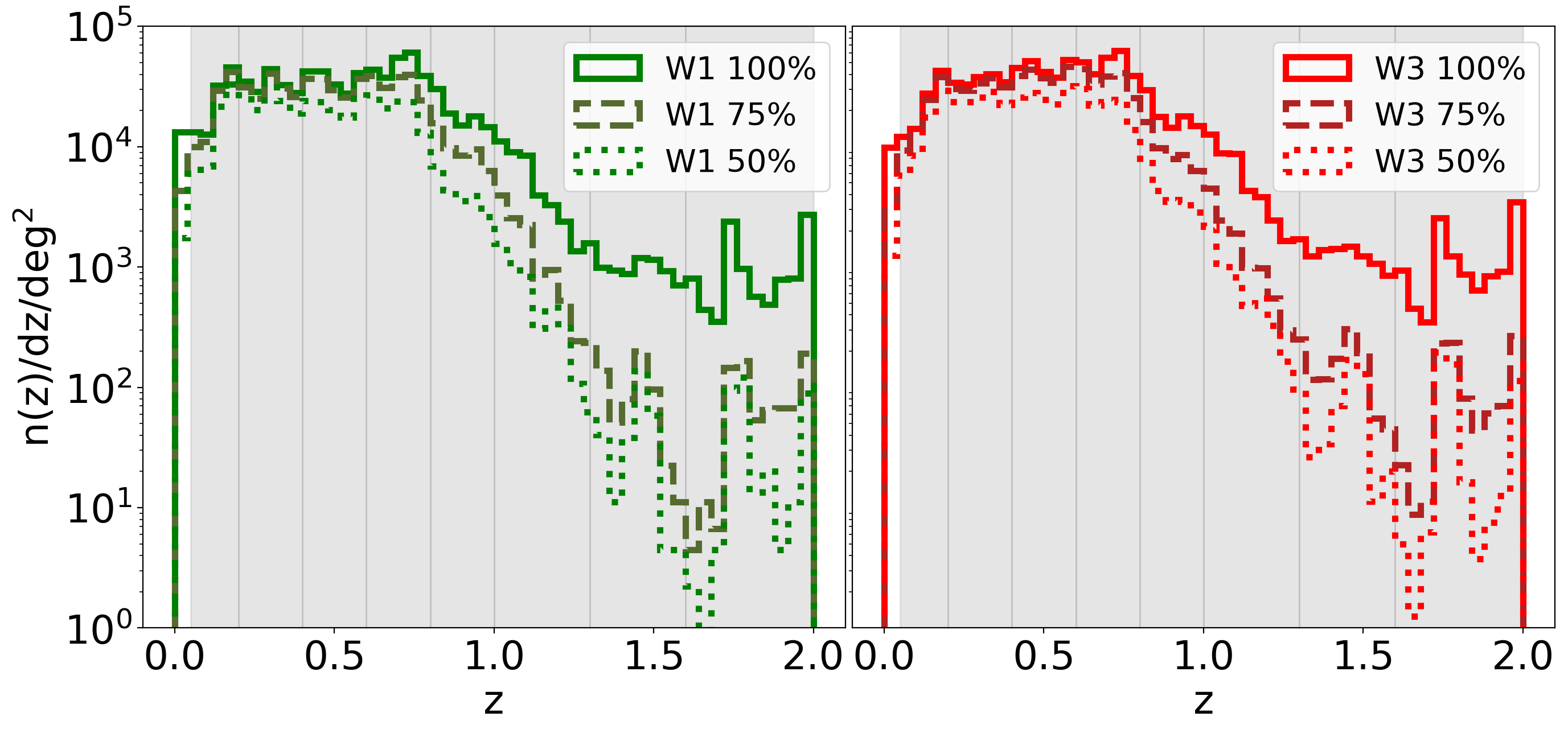}}%
\caption{The redshift distributions in the PAUS W1 (left) and W3 (right) fields. Note the $y$-axis in this case shows the logarithm of the number of galaxies to emphasize differences in the tail of the distribution. The thick solid line histogram shows the distribution for all galaxies brighter than $i_{\textrm{AB}}=23$.  The dashed line shows a cut on quality factor which retains the 75 per cent of galaxies with the best photometric redshifts and the dotted line shows the distribution for the best 50 per cent of redshifts. These cuts have the biggest impact on the number of galaxies in the high redshift tail.} %
    \label{fig:dndz_qz}%
\end{figure}

\begin{figure*}
\centering
    {\includegraphics[width=0.7\textwidth]{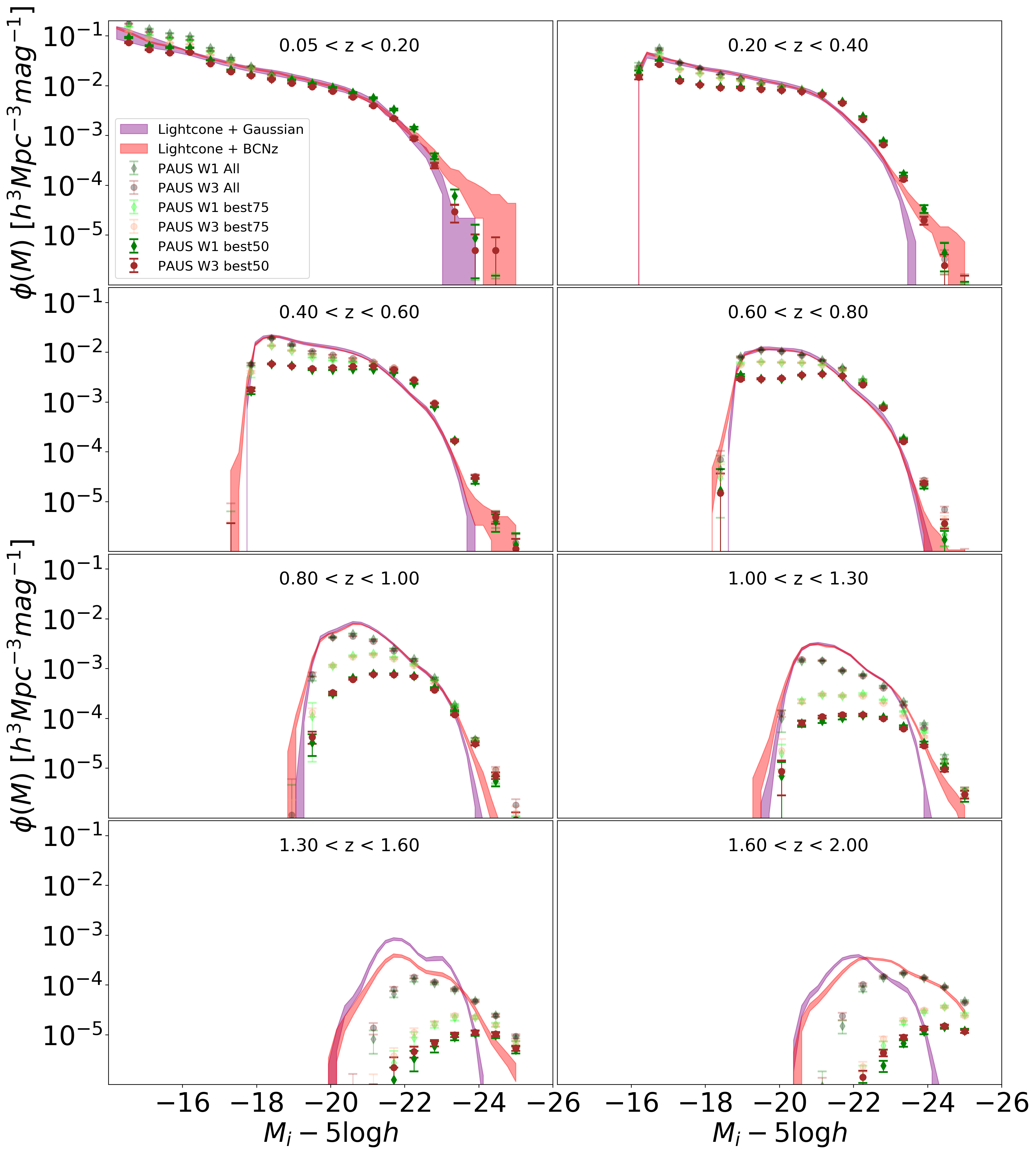}}%
\caption{The $i-$band galaxy luminosity function for multiple redshift bins ranging from $z = 0.05$ to $z = 2.00$. The LF is estimated for different sub-samples of galaxies with $i_{\textrm{AB}}=23$: all galaxies, the best 75 per cent of photometric redshifts as ranked by the quality factor and the best 50 per cent. The LFs estimated from the lightcone are also shown, for all galaxies, but with different models for the errors in the photometric redshifts: Gaussian error (purple), \texttt{BCNZ}2-like errors, including outliers (red).}%
    \label{fig:LF_best75}%
\end{figure*}

\subsection{Completeness and redshift quality}
% xxx
% At higher redshift ($z > 1.2$), the observed luminosity function from PAUS overestimates the bright end and underestimates the faint end compared to the \GALFORM predictions. As the \GALFORM lightcone mock catalogue is built specifically for the PAUS survey by introducing the same survey selection function, the discrepancies could result from the limitations in the observational data (e.g., photometric redshift uncertainties or photometry uncertainties) or differences in the underlying assumptions about galaxy formation processes in the model. The comparison between the two \GALFORM mock luminosity functions shows that the \texttt{BCNZ2}-based photometric errors lead to reduced discrepancies with the observation. We are able to reproduce the excess bright end at the two highest redshift bins. However with the improved agreement, the observation underestimates the faint-end of the luminosity function. PAUS is a imaging survey with a complete sample of galaxies down to the survey limit ($I_{AB} < 23$). Hence, the disagreement should not arise from the incompleteness of the sample at the faint end.  

We now return to the LF of the overall galaxy sample and consider how the redshift completeness and various systematic effects may influence our estimates. 

One immediate factor to take into account is any incompleteness due to the requirements on the observations of a galaxy for a photometric redshift to be estimated. A galaxy is required to be imaged in a large number of the NB filters (at least 30 filters out of the total of 40) before a photometric redshift is estimated. This was more of an issue when the mosaic of the camera pointings was still being built up to ensure that each field is visited multiple times, gaining exposure with a different tray of NB filters each time. At first sight, such an effect is seen in the number counts of various PAUS samples plotted in Fig. 2 of \cite{Manzoni2024}, where the number counts of galaxies with more than 30 NB detections were approximately $90$ per cent of the full galaxy sample. This implied that the LF normalisation should be adjusted upward by a factor of 1/0.9 to take this into account. However, a further investigation by us shows that this reduction was primarily due to a discrepancy in the effective survey area used and not a deficiency in the galaxy sample itself. Hence, when the normalisation of the LF is based on the correct effective  area, no further correction (e.g. the previously used  assumption of 1/0.9 factor) is required for photometric redshift completeness (as determined by the number of NB filters the galaxy has been imaged in).

% After performing star-galaxy separation, the counts of objects imaged in at least 30 narrow band filters was around 0.90 times the total galaxy count at the time the PAUS data was analysed for Manzoni et~al. If this factor had persisted, it would have been necessary to revise the normalisation of our LF estimates upwards by $1/0.9$. Subsequently, this factor is essentially 1, with all objects that are retained as galaxies after star-galaxy separation being imaged in 40 narrow band filters. 

Next we investigate the distribution and mean value of $V/V_{\textrm{max}}$ in each redshift shell. The $V/V_{\textrm{max}}$ for each galaxy is computed as following: $V$ is the volume covered by the redshift where the galaxy is located, with the volume covered by the lower limit of the redshift bin subtracted. $V_\textrm{max}$ is the volume covered by the maximum redshift that the galaxy appears to have the same apparent magnitude as the survey limit ($i_\textrm{AB} = 23$ in this case), with the volumce covered by the lower limit of the redshift bin substracted. This $V_\textrm{max}$ is the same volume used for weighting the LF. Hence, Fig.~\ref{fig:VVmax_all} shows the distribution of the $V/V_\textrm{max}$ of galaxies within each redshift bin. If a galaxy sample is a fair sample of the Universe, the mean value of $\langle V/V_{\textrm{max}} \rangle = 0.5  \pm 1/\sqrt{12 N}$ where $N$ is the total number of galaxies in the sample \citep{Peacock1999}. This error ($\pm 1/\sqrt{12 N}$) is for a uniform random distribution of points. For $20\,000$ galaxies, this error corresponds to $2 \times 10^{-3}$. The mean values of $V/V_{\textrm{max}}$ quoted in Fig.~\ref{fig:VVmax_all} differ from 0.5 by more than the size of the random error. This could indicate large-scale structure effects. The W3 field is just under twice the size of the W1 field. Fig.~\ref{fig:VVmax_all} shows large differences in the mean $V/V_{\textrm{max}}$ values returned for these fields. However, the estimates of the LFs from the W1 and W3 fields agree quite well with each other. 

Another issue affecting the $V/V_{\textrm{max}}$ distributions could be errors in the $k$-correction. These could lead to gradients in the $V/V_{\textrm{max}}$ distribution. However, the $k$-correction does not affect the $V_{\textrm{max}}$ value for most galaxies, since the maximum volume is simply the volume of the redshift slice used to measure the luminosity function. 

Finally, $V/V_{\textrm{max}}$ could be affected by evolution in the population of the galaxy. However, we expect this to be small as the redshift shells used to the measure the LF are small, and the LF does not evolve substantially between shells in the \texttt{GALFORM} model. 

Another consideration is the impact of what are believed to be low quality photometric redshifts on the recovered LFs. We have already argued that outliers affect the shape of the bright end of the LF at high redshift, by applying different scenarios for photometric redshift errors to the mock galaxies. We can isolate low quality photometric redshifts in the observations by applying cuts on the quality factor (equation 16 in \citealt{Eriksen2019}) and recomputing the LF. First, we look at the impact of making cuts on the quality factor on the redshift distribution in Fig.~\ref{fig:dndz_qz}.  The cuts on quality factor have a larger effect at high redshift. This agrees with the impression drawn from the photometric redshift versus spectroscopic redshift plot in Fig.~\ref{fig:specz_photoz_BCNz}, which shows a big increase in the proportion of outliers above a redshift of $\sim 1.2$. It is clear from the impact of a quality factor cut on the form of the redshift distribution that a simple revision to the amplitude of the LF will not be able to adjust for the change of the recovered LF (this would require a global shift in the redshift distribution, rather than a redshift-dependent change). 

The impact on the recovered LF of making cuts to remove low quality photometric redshifts is shown in Fig.~\ref{fig:LF_best75}. Removing the worst 25 per cent or 50 per cent of photometric redshifts leads to substantial changes in the estimated LFs, particularly at high redshifts. The largest change is at the faint end of the LF, as these galaxies are the faintest ones at a given redshift with the lowest signal to noise ratio and so are more prone to be affected by worse redshift estimates. There is a more modest reduction in the LF at the bright end. For the cut to remove the 25 per cent of galaxies with the worst photometric redshifts, the differences in the recovered LFs become significant from $z \sim 0.8$ and higher redshifts. The effect becomes greater when we further remove the worst 50 per cent of galaxies. Intriguingly, the $V/V_{\textrm{max}}$ distributions do not become more uniform after applying a cut on redshift quality. 

% \begin{figure}
% \centering
%     {\includegraphics[width=0.47\textwidth]{./figs/PAUS/LC_PAUS_W1_W3_LFs_z14_evo.png}}%
% \caption{The evolution of the $i$-band luminosity function from $z \sim 0.1$ to $z \sim 1.9$.}%
%     \label{fig:LC_PAUS_W1_W3_evo}%
% \end{figure}  

% \begin{figure}
% \centering
% {\includegraphics[width=0.47\textwidth]{./figs/PAUS/VVmax_distribution.png}}%
% \caption{The $V/V_{max}$ distribution for the galaxies in all the redshift bins used for the LFs. The gray, red, and blue histogram represents the lightcone mock catalogue, PAUS W1, and PAUS W3, respectively.}%
% \label{fig:VVmax_all}%
% \end{figure}  

\section{Sensitivity of the luminosity function to model parameters}
\label{sec:GALFORM variants}

\begin{figure}
\centering
    {\includegraphics[width=0.49\textwidth]{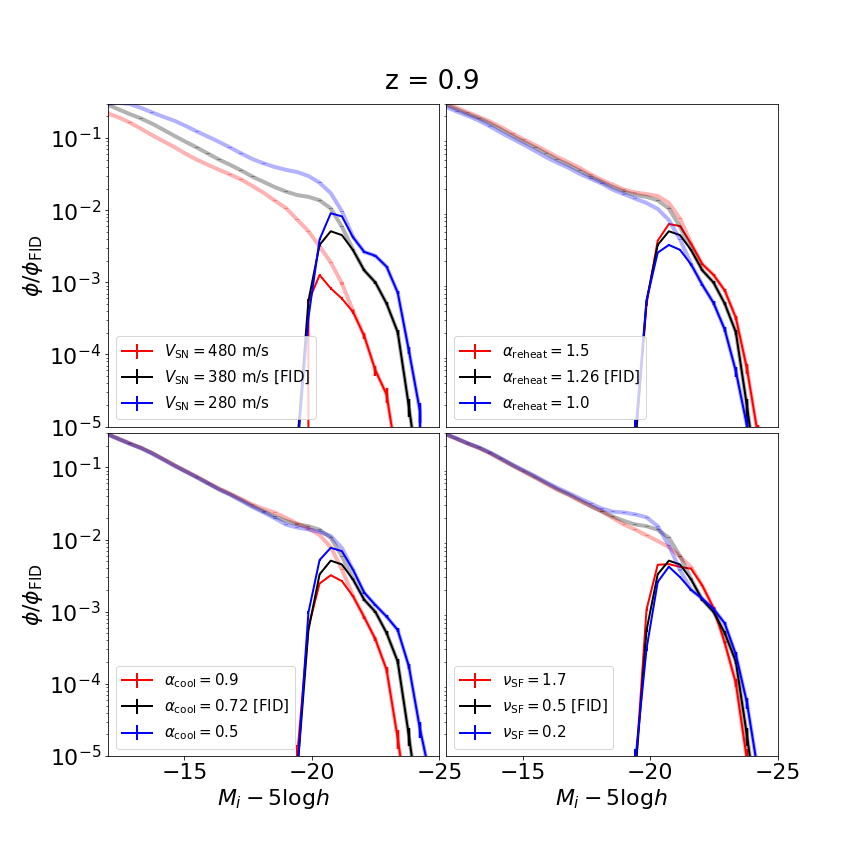}}%
\caption{The predicted $i$-band galaxy luminosity function with different model parameter variations at the redshift bin $z = 0.9$. The fiducial model is shown using black lines, whereas the higher and lower model values are plotted as red and blue lines, respectively. The true \texttt{GALFORM} LF is shown with fainter lines, while the LF with the $i_{\rm AB}=23$ selection function is plotted with darker lines. Each panel shows different the results from varying different model parameters, as labelled at the lower left of each panel.} %
    \label{fig:LF_GALFORM_variants}%
\end{figure}

\begin{figure}
\centering
    {\includegraphics[width=0.49\textwidth]{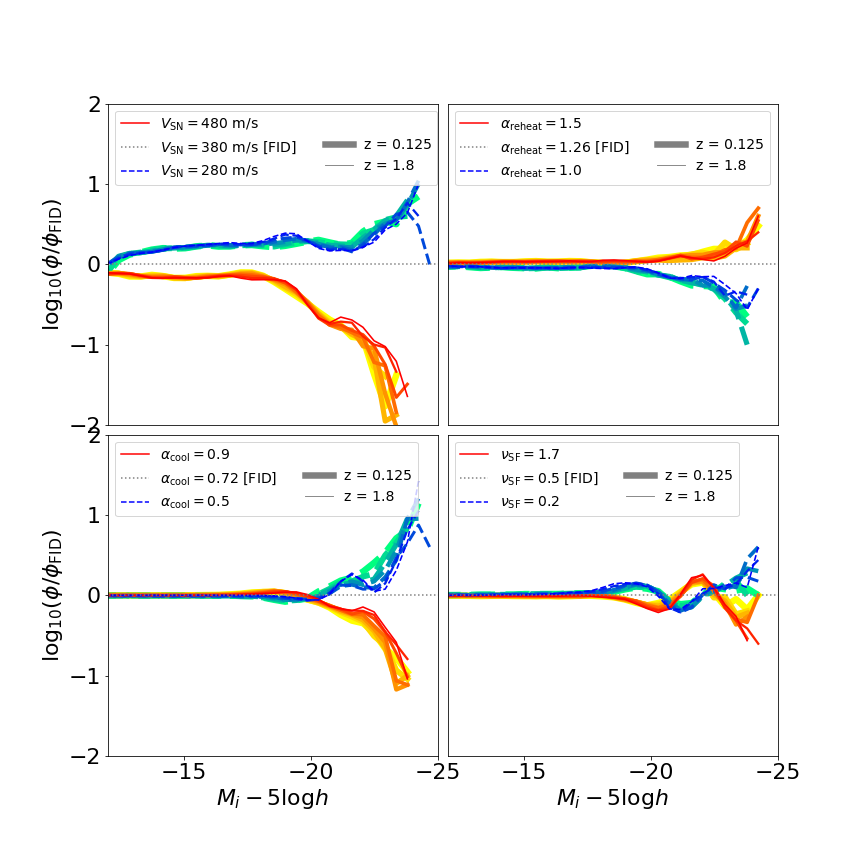}}%
\caption{The fractional change of the LF variants relative to the fiducial model. The horizontal dotted line indicates no deviation from the fiducial prediction. Red-ish and blue-ish colours represent the predictions with higher and lower model parameter variations, respectively. The width of each line represents the redshift bin the LF is measured in, with thicker lines for low redshift bins and thinner lines for higher redshift bins.}%
    \label{fig:Feedback_strength_nocut}%
\end{figure}

\begin{figure}
\centering
    {\includegraphics[width=0.49\textwidth]{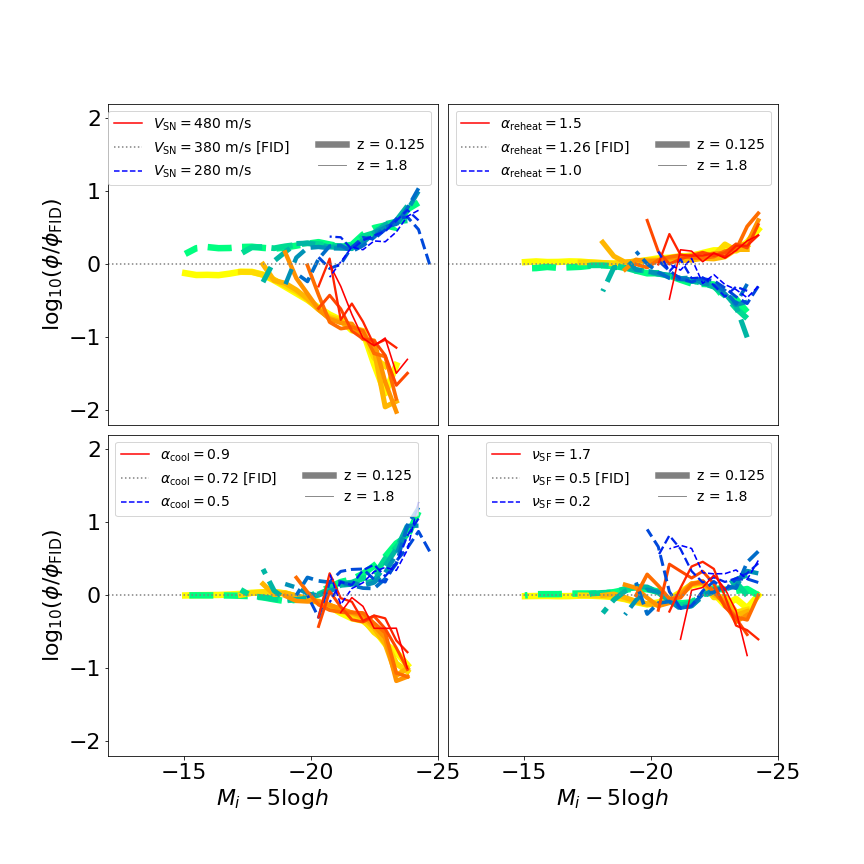}}%
\caption{The same description as Fig.~\ref{fig:Feedback_strength_nocut}, but with the $i_{\rm AB} = 23$ selection function applied.} %
    \label{fig:Feedback_strength}%
\end{figure}

In this section, we explore the sensitivity of the predicted galaxy luminosity function to variations of the \texttt{GALFORM} model parameters and see how these  differences compare to systematic effects in the recovered LF. This analysis follows the approach of \citet{Manzoni2024}. We alter a subset of processes that are known to have a large effect on the galaxy LF \citep{Piotr2020}. These processes govern the star formation activity and hence affect the intrinsic luminosities of galaxies, which transforms directly into changes in the resulting luminosity functions. These parameters, listed in Table~3, include the strength of the supernova (SNe) driven winds, the timescale for SNe heated gas to be reincorporated into the hot halo, the efficiency of heating by active galactic nuclei (AGN) and the subsequent suppression of gas cooling, and the time scale for quiescent star formation. 

\begin{table}
\centering
\begin{tabular}{ cccc } 
\hline
Parameters & low & fiducial & high\\
\hline
$V_{\rm SN}\ (\rm km \ s^{-1})$ &  280 & 380 & 480\\ 
$\alpha_{\rm reheat}$ & 1.00 & 1.26 & 1.50 \\
$\alpha_{\rm cool}$ & 0.50 & 0.72 & 0.90 \\
$\nu_{\rm SF}$ & 0.20 & 0.50 & 1.70\\
\hline
\end{tabular}
\caption{The parameter values in the variant models. The first columns provides the parameter names. The third column gives the fiducial value of the parameter, whereas the second and fourth columns give the low and high values considered, respectively.}
\label{Tab:GALFORM_variants}
\end{table}

Four model parameters are changed in this exercise, one at a time keeping other parameters fixed, resulting in eight variant models. The parameter values are listed in Table~\ref{Tab:GALFORM_variants}: (i) the pivot velocity that controls the mass loading of SNe driven winds, $V_{\rm SN}$ (Equation 10 in \citealt{Lacey2016}), with higher values resulting in larger mass ejection rates from more massive halos (ii) the timescale for gas heated by SNe to be reincorporated into the hot gas halo, which is inversely proportional to $\alpha_{\rm reheat}$ (Equation 11 in \citealt{Lacey2016}), with larger values giving shorter reincorporation times (iii) the star formation efficiency factor, $\nu_{\rm SF}$, (Equation 7 of \citealt{Lacey2016})\footnote{The $\nu_{\rm SF}$ variants listed in Table~3 correspond to the full range suggested by observations of local star forming galaxy \citep{Blitz-Rosolowsky06}.}; and (iv) the factor which determines the halos mass in which AGN heating  starts to prevent to cooling of gas, $\alpha_{\rm cool}$ (Equation 12 in \citealt{Lacey2016}). For each parameter, we consider two alternative values bracketing the value used in the fiducial model, resulting in eight variant models in total. The parameter values used are listed in Table~\ref{Tab:GALFORM_variants}. Note that we do not change any other model paramaters to compensate for the changes introduced to the LF by varying the selected parameter, so the variant models do not necessarily reproduce the calibration data with the same level of agreement as the fiducial model.

Fig.~\ref{fig:LF_GALFORM_variants} shows the predicted rest-frame $i$-band LF at $z = 0.9$ for each parameter variation. The red, black, and green lines correspond to the higher, fiducial, and lower parameter values, respectively. We plot the \texttt{GALFORM} LF without the $i_{AB} = 23$ cut using fainter lines, while we plot the LFs with this magnitude cut applied on top with darker shades. The effect of varying SNe feedback strength results in the largest changes at the bright end of the LF (top-left panel of Fig.~\ref{fig:LF_GALFORM_variants}).

Fig.~\ref{fig:Feedback_strength_nocut} shows the fractional change relative to the fiducial model, expressed as $\log_{10}(\Phi/\Phi_{\rm FID})$. The horizontal dotted line indicates no deviation from the fiducial prediction. Across all redshift bins, the strongest impact arises from the parameter controlling driven winds, $V_{\rm SN}$. Lowering $V_{\rm SN}$ increases the number density of galaxies across all magnitudes, with a stronger effect at the bright end. This corresponds to an increase of $\sim 0.3$ dex overall and up to $\sim 1$ dex at the bright end. Conversely, increasing $V_{\rm SN}$ suppresses the LF. A similar trend is seen for $\alpha_{\rm reheat}$, although with a smaller change amplitude, particularly at the faint end, and a change of approximately 0.5 dex at the bright end. Variations in $\alpha_{\rm cool}$ primarily affect the bright end through AGN feedback. Higher $\alpha_{\rm cool}$ values shift the onset of AGN heating to lower halo masses, suppressing cooling more efficiently, and reduce the abundance of bright galaxies. In contrast, changes in the star formation efficiently parameter $\nu_{\rm SF}$ produce a more complex response. Increasing $\nu_{\rm SF}$ boosts star formation in galaxies around the characteristic magnitude $M^*$ (the "knee" of the LF), while slightly reducing both faint- and bright-end number densities. Lower values of $\nu_{\rm SF}$ result in the opposite behaviour. Overall, the redshift dependence of these trends is weak, suggesting that the qualitative behaviour of each parameter variation remains similar as a function of redshift.

Fig.~\ref{fig:Feedback_strength} shows the same fractional differences after applying the $i_{\rm AB} = 23$ selection function to the sample used in Fig.~\ref{fig:Feedback_strength_nocut}. For most parameters ($V_{\rm SN}$, $\alpha_{\rm reheat}$, and $\alpha_{\rm cool}$), the amplitudes remain largely unchanged once the selection effect is included. The faint-end deviations are suppressed by the magnitude limit, as discussed in the previous section, and the parameter variations primarily appear as a rescaling of the LF at bright magnitudes. However, the case of $\nu_{\rm SF}$ is more sensitive to selection. Variations in star formation efficiency alter the balance between red and blue galaxy populations, making the predicted LF more sensitive to the transition between the complete and incomplete regimes of the survey, particularly near the LF knee at higher redshifts.

\begin{figure}
\centering
    {\includegraphics[width=0.49\textwidth]{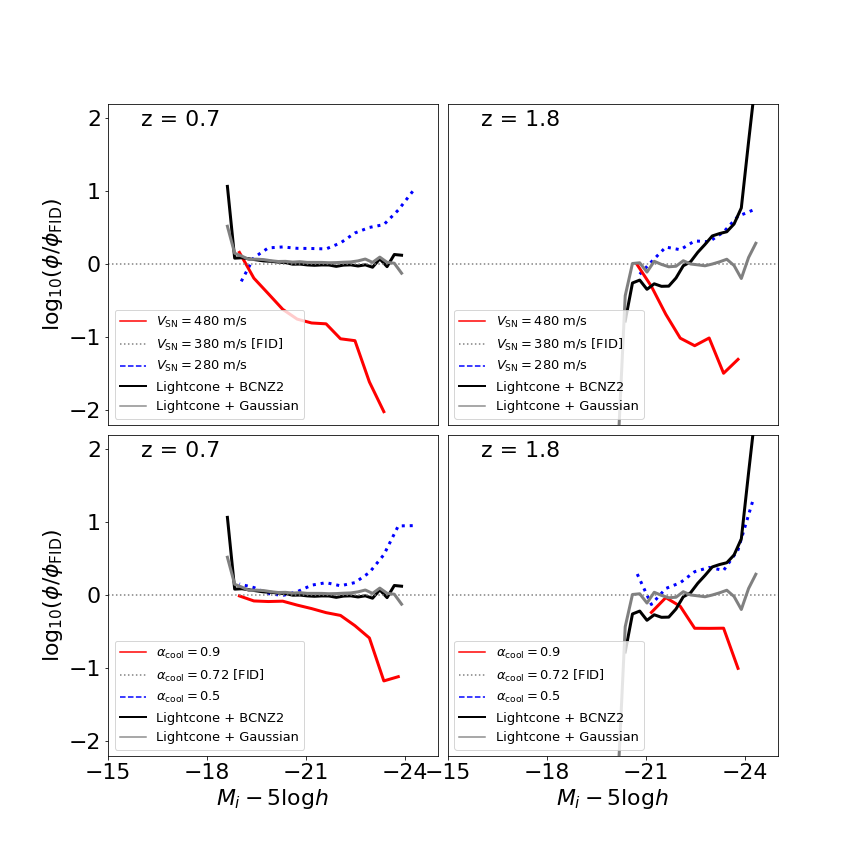}}%
\caption{A comparison between the $i_{AB}=23$ selected \texttt{GALFORM} LF and the lightcone LF. We plotted the lightcone LF with two variants of photometric redshift errors applied: grey lines for Gaussian errors and black lines for \texttt{BCNz2} errors. The left and right panels show a comparison at $z = 0.7$ and $z = 1.8$, respectively.}%
    \label{fig:Feedback_BCNZ}%
\end{figure}

Next, we investigate whether variations of the model parameters could be recovered given the overprediction at the bright end of the LF after applying the BCNz2-like photometric redshift errors. We find that lower values of $V_{\rm SN}$ and $\alpha_{\rm cool}$ reduce the discrepancy at the bright end in the redshift bin $z = 1.8$, as shown in the right panel of Fig.~\ref{fig:Feedback_BCNZ}. However, applying the same parameter variations leads to an overestimation of the LF at lower redshift (left panel of Fig.~\ref{fig:Feedback_BCNZ}). Moreover, none of the parameter variations reproduce galaxies brighter then $M_i - 5\log h = -24.5$, which is evident when BCNz2 errors are included (see Fig.~\ref{fig:LC_PAUS_W1_W3}). This indicates that the change in the shape of the high redshift ($z>1$) LF at the bright end driven primarily by photometric redshift outliers and this restricts the ability of the PAUS measurements to distinguish between different galaxy formation models at these redshifts.

\section{Summary and conclusions}

We have presented a measurement of the rest-frame $i$-band galaxy luminosity function (LF) using over one million galaxies from the Physics of the Accelerating Universe Survey (PAUS; \citealt{Padilla2019}), reaching up to a redshift of $z \sim 2$. The combination of CFHTLenS broad-band imaging and PAUS narrow-band photometry enables photometric redshift measurements with unprecedented accuracy, reaching $\sigma_{68} (\Delta z) = 0.019$ for galaxies with $i_{\textrm{AB}} < 23$. This precision, which is intermediate between that typically achieved with a small number of broad-band filters and spectroscopic accuracy, allows us to estimate the LF across more than 60 per cent of the age of the Universe, with reduced uncertainty due to redshift errors and sample variance (due to analysing a combined area of more than 30~deg$^2$).

We used a synthetic galaxy catalogue constructed by \cite{Manzoni2024} using the semi-analytical galaxy formation model \GALFORM applied to the high-resolution Planck Millennium N-body simulation \citep{Baugh2019}, producing a mock PAUS survey with selection criteria and photometric characteristics consistent with the actual survey. This allowed us to quantify the impact of selection effects, magnitude and redshift errors, and incompleteness on the recovered LF. A key conclusion reached from analysing the data alongside a realistic mock is that we can find out over what magnitude range the estimate of the LF becomes incomplete; moreover, we can still use the recovered LF at these magnitudes to constrain the galaxy formation model. In addition to constraining the form of the LF, the colour distribution of galaxies also shapes the LF over the range of magnitudes for which the LF is incomplete.   

We developed a machine learning method to estimate $k$-corrections based on random forest regression (RFR), using $ugriz$ photometry and the redshift as inputs. The method is trained using the mock catalogue, which provides the exact $k$-correction at the redshift of observation, for a population of galaxies with varied (non-parametric) star formation histories.  This approach accurately recovers the rest-frame $i$-band absolute magnitudes up to $z \le 1$, although it introduces a bias of up to $\sim0.2$ magnitudes at higher redshifts due to the increased scatter in photometric redshifts and intrinsic colour variations.

We measured the galaxy LF using the $1/V_{\textrm{max}}$ estimator in narrow redshift bins, incorporating realistic photometry and photo-z errors. Comparisons between the observed LF from the PAUS W1 and W3 fields and predictions from the GALFORM mock showed good agreement at $z < 1$, with larger discrepancies emerging at higher redshifts. The $i$-band selection introduces a turnover at faint magnitudes, which depends upon the underlying LF and the colour distribution of galaxies. These selection effects can be accurately modelled through the mock, so, although the measurement of the LF is incomplete at these magnitudes, it still contains useful information that can be used to constrain galaxy formation models. At $z \ge 1.3$, the inclusion of photometric redshift outliers in the mock catalogues is essential to reproduce the observed flattening of the LF at the bright end. This demonstrates that photo-z outliers can significantly distort the luminosity distribution, especially where the galaxy number density declines steeply with luminosity.

Weak gravitational lensing is unlikely to have an important effect on our estimate of the LF. The LF at the highest redshift, $z\sim 2$ is likely to be most affected by lensing. Several calculations have argued that the fluctuations in the magnification factor induced by weak lensing are at the level of $5-10$ per cent \citep{DA2014, Lee2016,Max:2021}.
At $z\sim2$, the recovered LF is much brighter than the break of the underlying LF, so the estimates are part of the exponential break. The impact of magnification bias is smaller than other uncertainties, such as the error in the $k$-correction or the photometric redshifts. 

We further investigated the LF of red and blue galaxy populations separately, dividing galaxies based on the observed $g-r$ colour (using the condition set out in \citealt{Manzoni2024}). As expected, red galaxies dominate the bright end of the LF, exhibiting a pronounced exponential cut-off, whilst blue galaxies dominate at fainter magnitudes and display a steeper faint-end slope. This bimodality persists over the full redshift range. The evolution of the LF break luminosity ($L_*$) is more prominent for blue galaxies, shifting to brighter magnitudes with increasing redshift. The mock catalogue over-predicts the number of faint galaxies in both populations, suggesting potential issues in modelling feedback processes or the modelling of gas cooling in satellites (see, for example, \citealt{Font2008}), particularly for low-mass galaxies.

To assess the completeness of the sample, we examined the $V/V_{\textrm{max}}$ statistic in each redshift bin. Deviations from the expected value of 0.5 suggest that large-scale structure, selection effects or errors in the $k$-correction--may introduce incompleteness. However, the consistency between the W1 and W3 field estimates supports the robustness of our LF measurements with respect to LSS. We further examined the impact of photometric redshift quality by applying cuts on the photometric redshift quality factor. Removing the worst 25–50 per cent of redshifts has a substantial effect on the LF at high redshifts, particularly at the faint end.

Our results are broadly consistent with previous measurements of the LF from both spectroscopic and photometric surveys, including those by \cite{Ilbert2005, Ramos2011} and \cite{Driver2012}, though differences in depth and solid angle coverage naturally lead to variations in the accessible magnitude and redshift ranges. Compared to these surveys, PAUS provides a unique balance of depth, area, and redshift precision.

This work underscores the scientific potential of photometric surveys with high accuracy photometric redshifts such as PAUS. By combining deep narrow-band imaging, robust mock catalogues, and machine learning techniques, we demonstrate that photometric redshifts can be used to recover galaxy luminosity functions with precision approaching that of spectroscopic surveys. The success of our methodology, testing a galaxy formation model against the observational estimate of the LF, points to a way to test the models over a wider range of luminosities than those for which the LF is formally complete. 

However, our analysis also highlights several areas where improvements are needed. The impact of photometric redshift outliers at $z > 1.3$ introduces uncertainties that are difficult to correct for without spectroscopic calibration. The \texttt{GALFORM} model overpredicts the abundance of faint galaxies, pointing to an overproduction of low-luminosity systems or an insufficient treatment of feedback in the model. Further, the $k$-correction method, whilst effective at moderate redshifts, introduces biases at higher redshifts that may be mitigated through the use of extended training datasets improving the wavelength coverage of the photometry. 

In summary, the PAUS $i$-band luminosity function offers a new benchmark for measuring galaxy evolution with photometric redshifts. With its unique dataset and careful treatment of systematics, this work paves the way for precision studies of the galaxy population across cosmic time using current and upcoming photometric surveys.

\section*{acknowledgements}
The authors would like to thank Peder Norberg, Shaun Cole, and Nicha Leethochawalit for useful conversations about this project.
%SK is supported by the National Astronomical Research Institute of Thailand under the Office of the Civil Service Commission (OCSC) studentship from the Royal Thai Government
SK is supported by the Fundamental Fund of Thailand Science Research and Innovation (TSRI) through the National Astronomical Research Institute of Thailand (Public Organisation) (FFB680072/0269). 
CMB acknowledges support from STFC through ST/X001075/1. 
GM acknowledges the support of the SKL of Displays and Opto-Electronics (project 2023 SKL ITC-PSKL12EG02 R1X27).
DNG and H. Hoekstra acknowledge support from the European Research Council (ERC) under the European Union’s Horizon 2020 research and innovation program with Grant agreement No. 101053992.
H. Hildebrandt is supported by a DFG Heisenberg grant (Hi 1495/5-1), the DFG Collaborative Research Center SFB1491, an ERC Consolidator Grant (No. 770935), and the DLR project 50QE2305.
EG acknowledge support from the Spanish Ministerio de Ciencia e Innovacion (MICINN), project PID2021-128989NB.
JGB acknowledges support from the Spanish Research Project PID2021-123012NB-C43 [MICINN-FEDER], and the Centro de Excelencia Severo Ochoa Program CEX2020-001007-S at IFT.
FJC acknowledges support from the Spanish Plan Nacional project PID2022-141079NB-C31.
ME acknowledges funding by MCIN with funding from European Union NextGenerationEU (PRTR-C17.I1) and by Generalitat de Catalunya.
This work used the DiRAC@Durham facility managed by the Institute for Computational Cosmology on behalf of the STFC DiRAC HPC Facility (www.dirac.ac.uk). The equipment was funded by BEIS capital funding via STFC capital grants ST/K00042X/1, ST/P002293/1, ST/R002371/1, and ST/S002502/1, Durham University and STFC operations grant ST/R000832/1. DiRAC is part of the National e-Infrastructure. The PAU Survey is partially supported by MINECO under grants CSD2007-00060, AYA2015-71825, ESP2017-89838, PGC2018-094773, PGC2018- 102021, PID2019-111317GB, SEV-2016-0588, SEV-2016-0597, MDM-2015-0509 and Juan de la Cierva fellowship, and LACEGAL and EWC Marie Sklodowska-Curie grants 101086388 and 776247 from the EU Horizon 2020 Program. 
Funding for PAUS has also been provided by Durham University (via the ERC StG DEGAS-259586), ETH Zurich, Leiden University (via ERC StG ADULT-279396 and Netherlands Organisation for Scientific Research (NWO) Vici grant 639.043.512), University College London and from the European Union's Horizon 2020 research and innovation programme under the grant agreement No 776247 EWC. The PAU data center is hosted by the Port d'Informaci\'o Cient\'ifica (PIC), maintained through a collaboration of CIEMAT and IFAE, with additional support from Universitat Aut\`onoma de Barcelona and ERDF.
This work has also used CosmoHub, developed by PIC (maintained by IFAE and CIEMAT) in collaboration with ICE-CSIC. It received funding from the Spanish government (grant EQC2021-007479-P funded by MCIN/AEI/10.13039/501100011033), the EU NextGeneration/PRTR (PRTR-C17.I1), and the Generalitat de Catalunya.

\section*{Data availability}
The data used in this study are stored at the Durham COSMA facility (\texttt{GALFORM} mock) and at the Barcelona Port d’Informació Científica (PIC; PAUS observations). The \texttt{GALFORM} mock is available from GM whilst the observational data is available in the PAUS master catalog 1.0 (https://cosmohub.pic.es/catalogs/319).

%\end{acknowledgements}
%
\bibliographystyle{mnras}
\bibliography{suttikoon_ref}

\begin{appendix}

\section{CFHTLENS \ \  \texorpdfstring{\lowercase{$ugrz$}-}- band luminosity functions}

Here we present the results for the LFs in the other broad bands used to image the CFHTLenS W1 and W3 fields. The shape of the turnover at the faint end varies differently with redshift as the band changes, compared with the $i$-band results and the other bands. The LF is estimated in each case using the $V_{\textrm{max}}$ weights computed using the $i$-band selection. 

\begin{figure*}
\centering
    {\includegraphics[width=0.9\textwidth]{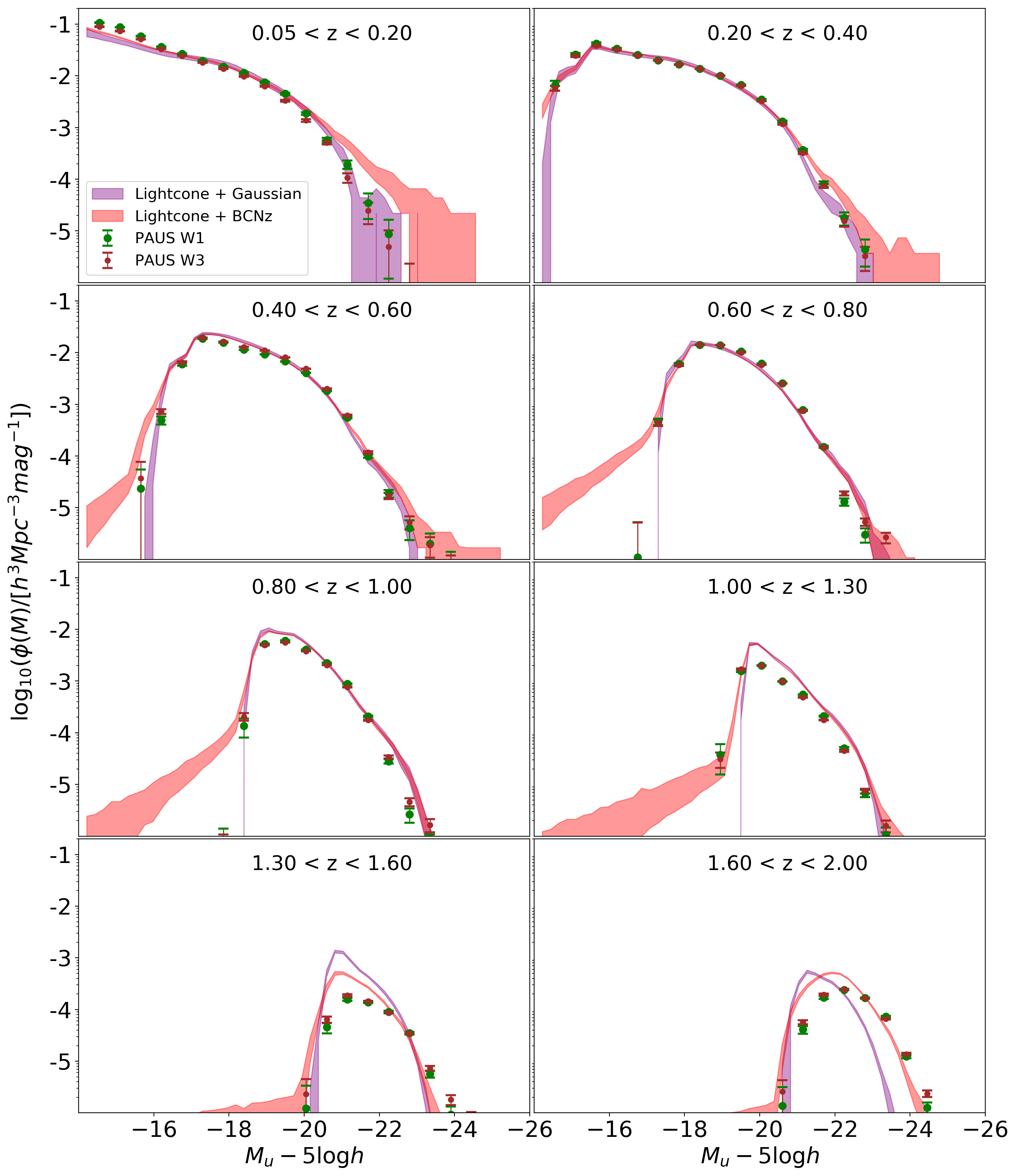}}%
\caption{The same description as Figure \ref{fig:LC_PAUS_W1_W3}, but for the $u$-band luminosity function.}%
    \label{fig:LC_PAUS_W1_W3_uband}%
\end{figure*}  

\begin{figure*}
\centering
    {\includegraphics[width=0.9\textwidth]{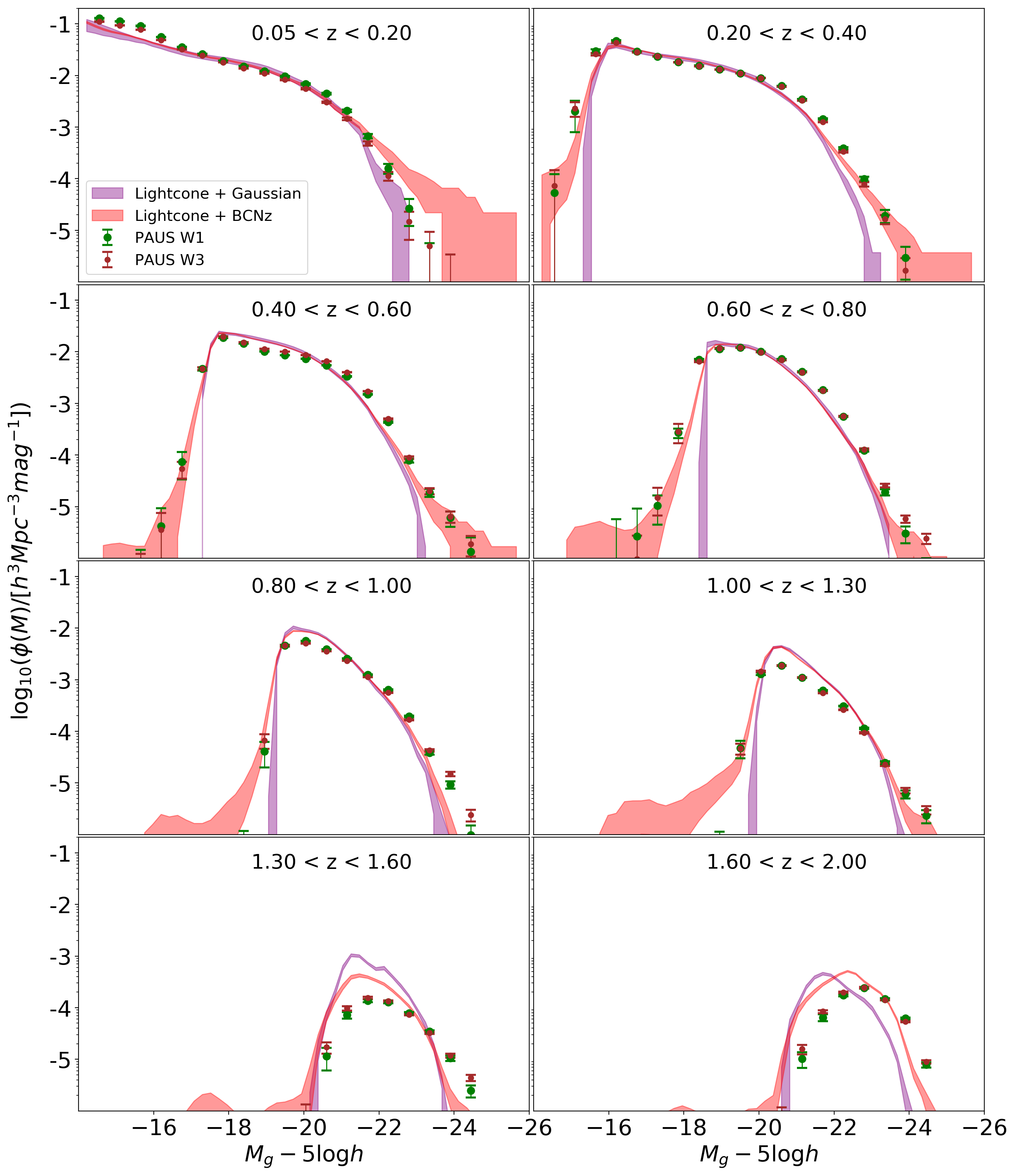}}%
\caption{The same description as Figure \ref{fig:LC_PAUS_W1_W3}, but for the $g$-band luminosity function.}%
    \label{fig:LC_PAUS_W1_W3_gband}%
\end{figure*}

\begin{figure*}
\centering
    {\includegraphics[width=0.9\textwidth]{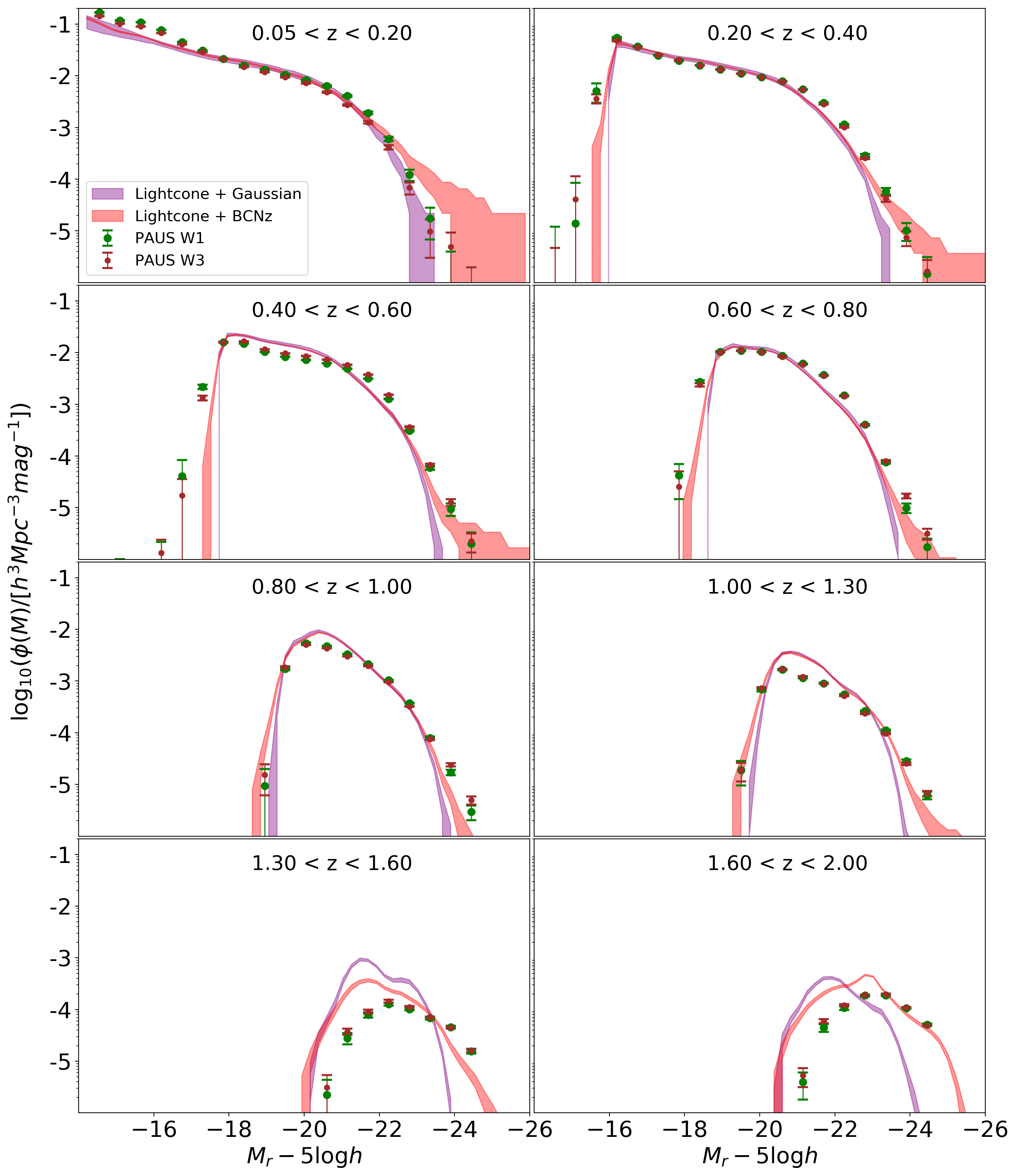}}%
\caption{The same description as Figure \ref{fig:LC_PAUS_W1_W3}, but for the $r$-band luminosity function.}%
    \label{fig:LC_PAUS_W1_W3_rband}%
\end{figure*}  

\begin{figure*}
\centering
    {\includegraphics[width=0.9\textwidth]{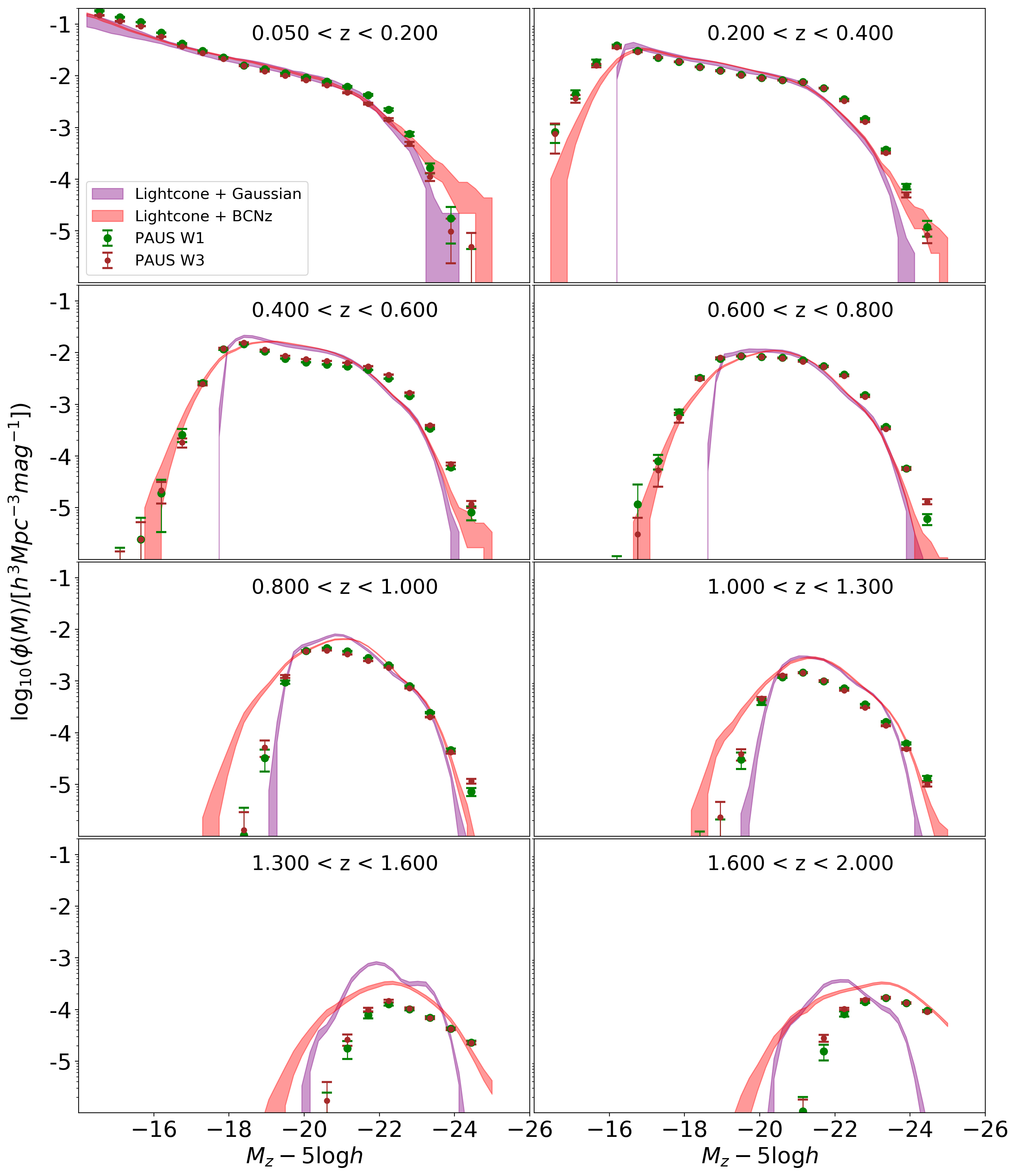}}%
\caption{The same description as Figure \ref{fig:LC_PAUS_W1_W3}, but for the $z$-band luminosity function.}%
    \label{fig:LC_PAUS_W1_W3_zband}%
\end{figure*}

\section{The $i$-band luminosity function tables}
Here we present the measured rest-frame $i$-band luminosity function in tabular form for redshifts up to $z = 1$ (i.e. the first five redshift bins of Fig~\ref{fig:LC_PAUS_W1_W3}). The columns are defined as follows: $M_i$ denotes the absolute magnitude bins; $\Phi_{\rm true}^{LC}$ is the luminosity function measured from the \texttt{GALFORM} lightcone mock catalogue without observational uncertainties;  $\Phi_{\rm Gau}^{LC}$ is the mock luminosity function after applying Gaussian-like photometric redshift errors;  $\Phi_{\rm BCNZ}^{LC}$ is the mock luminosity function after applying realistic PAUS-like photometric redshift errors:  $\Phi_{\rm W1}^{PAUS}$ is the luminosity function derived from the PAUS W1 field; and  $\Phi_{\rm W3}^{PAUS}$ is the luminosity function derived from the PAUS W3 field.

For readability in Tables~\ref{tab:LF_Tables_1} and \ref{tab:LF_Tables_2}, all luminosity function values have been multiplied by a factor of $10^3$. For each redshift bin, the magnitude limit above which luminosity function is considered complete (see Fig.~\ref{fig:GALFORM_LF_selection}) is indicated by a horizontal line in the table.

Tables for the remaining luminosity functions are provided as supplementary material.

\begin{landscape}
\begin{table}
    \centering
    \begin{tabular}{cccccccccccccccc}
    % \hline
    %    1& 2& 3& 4& 5& 6& 7& 8& 9&10&11&12&13&14&15&16\\
    \hline
          &      \multicolumn{5}{c}{0.05 < $z$ < 0.2} & \multicolumn{5}{c}{0.2 < $z$ < 0.4} & \multicolumn{5}{c}{0.4 < $z$ < 0.6}\\
    \cmidrule(lr){2-6}  \cmidrule(lr){7-11} \cmidrule(lr){12-16}
     $M_i$&$\Phi^{\rm{LC}}_{\rm{true}}$&$\Phi^{\rm{LC}}_{\rm{Gau}}$&$\Phi^{\rm{LC}}_{\rm{BCNZ}}$&$\Phi^{\rm{PAUS}}_{\rm{W1}}$&$\Phi^{\rm{PAUS}}_{\rm{W3}}$&$\Phi^{\rm{LC}}_{\rm{true}}$&$\Phi^{\rm{LC}}_{\rm{Gau}}$&$\Phi^{\rm{LC}}_{\rm{BCNZ}}$&$\Phi^{\rm{PAUS}}_{\rm{W1}}$&$\Phi^{\rm{PAUS}}_{\rm{W3}}$&$\Phi^{\rm{LC}}_{\rm{true}}$&$\Phi^{\rm{LC}}_{\rm{Gau}}$&$\Phi^{\rm{LC}}_{\rm{BCNZ}}$&$\Phi^{\rm{PAUS}}_{\rm{W1}}$&$\Phi^{\rm{PAUS}}_{\rm{W3}}$\\
\hline
     -26.10 & 0& 0 & \uncertainty{0.006}{+0.015}{-0.006} & 0 & 0&
     
     0& 0&  \uncertainty{0.001}{+0.003}{-0.001}&  0 &  0&
     
     0&  0&  \uncertainty{0.001}{+0.001}{-0.001}&  0& 0
     \\

     -25.55 & 0&0&\uncertainty{0.014}{+0.008}{-0.014}& 0 & 0&
              0& 0&\uncertainty{0.001}{+0.002}{-0.001}& 0 & 0&
              0& 0&\uncertainty{0.001}{+0.001}{-0.001}& \uncertainty{0.001}{+0.001}{-0.001}&\uncertainty{0.001}{+0.001}{-0.001}
              \\
     -25.00 & 0&0&\uncertainty{0.02}{+0.03}{-0.02}& 0 & 0 &
              0& 0&\uncertainty{0.004}{+0.003}{-0.004}& 0 & \uncertainty{0.001}{+0.001}{-0.001}&
              0& 0&\uncertainty{0.002}{+0.002}{-0.002}& \uncertainty{0.001}{+0.001}{-0.001}& \uncertainty{0.001}{+0.001}{-0.001}
              \\
     -24.45 & 0& 0&\uncertainty{0.04}{+0.03}{-0.04}& 0 & \uncertainty{0.005}{+0.004}{-0.004} &
              0& 0&\uncertainty{0.008}{+0.006}{-0.005}& \uncertainty{0.004}{+0.002}{-0.002} & \uncertainty{0.004}{+0.002}{-0.002} &
              0& 0&\uncertainty{0.003}{+0.003}{-0.002}& \uncertainty{0.005}{+0.001}{-0.001}& \uncertainty{0.005}{+0.001}{-0.001}  
              \\
     -23.90 & 0&\uncertainty{0.01}{+0.02}{-0.01}&\uncertainty{0.06}{+0.04}{-0.04}&\uncertainty{0.009}{+0.008}{-0.008}&\uncertainty{0.005}{+0.005}{-0.005}&
              0& 0&\uncertainty{0.02}{+0.01}{-0.01}& \uncertainty{0.034}{+0.006}{-0.006} & \uncertainty{0.021}{+0.004}{-0.004}&
              0& 0&\uncertainty{0.007}{+0.004}{-0.004}& \uncertainty{0.027}{+0.004}{-0.004}& \uncertainty{0.032}{+0.003}{-0.003}
              \\
     -23.35 &0.02&\uncertainty{0.02}{+0.01}{-0.02}&\uncertainty{0.15}{+0.05}{-0.06}&\uncertainty{0.06}{+0.02}{-0.02}&\uncertainty{0.03}{+0.01}{-0.01}&
            0.04&\uncertainty{0.04}{+0.01}{-0.01}&\uncertainty{0.09}{+0.01}{-0.01}&\uncertainty{0.17}{+0.01}{-0.01}& \uncertainty{0.14}{+0.01}{-0.01}&
            0.06&\uncertainty{0.07}{+0.01}{-0.01}&\uncertainty{0.07}{+0.01}{-0.01}&\uncertainty{0.17}{+0.01}{-0.01}&\uncertainty{0.174}{+0.006}{-0.006}
            \\
     -22.80 &0.24&\uncertainty{0.22}{+0.09}{-0.07}&\uncertainty{0.42}{+0.10}{-0.10}&\uncertainty{0.39}{+0.05}{-0.05}&\uncertainty{0.27}{+0.03}{-0.03}&
            0.26&\uncertainty{0.32}{0.04}{0.05}&\uncertainty{0.34}{+0.03}{-0.03}&\uncertainty{0.78}{+0.03}{-0.03}& \uncertainty{0.67}{+0.02}{-0.02}&
            0.42&\uncertainty{0.46}{+0.06}{-0.06}&\uncertainty{0.43}{+0.02}{-0.02}&\uncertainty{0.82}{+0.02}{-0.02}& \uncertainty{0.98}{+0.01}{-0.01}  
            \\
     -22.25 &0.91&\uncertainty{0.82}{+0.14}{-0.14}&\uncertainty{1.08}{+0.14}{-0.14}&\uncertainty{1.43}{+0.10}{-0.10}&\uncertainty{0.92}{+0.05}{-0.05}&
            1.05&\uncertainty{1.05}{+0.13}{-0.15}&\uncertainty{1.14}{+0.06}{-0.05}&\uncertainty{2.42}{+0.04}{-0.04}& \uncertainty{2.17}{+0.03}{-0.03}&
            1.18&\uncertainty{1.24}{+0.10}{-0.09}&\uncertainty{1.23}{+0.03}{-0.03}&\uncertainty{2.42}{+0.03}{-0.03}& \uncertainty{2.88}{+0.03}{-0.03}  
            \\
     -21.70 &2.49&\uncertainty{2.29}{+0.25}{-0.26}&\uncertainty{2.22}{+0.20}{-0.20}&\uncertainty{3.34}{+0.15}{-0.15}&\uncertainty{2.26}{+0.09}{-0.09}&
            2.81&\uncertainty{2.68}{+0.25}{-0.28}&\uncertainty{2.72}{+0.08}{-0.08}&\uncertainty{4.80}{+0.06}{-0.06}&\uncertainty{4.62}{+0.05}{-0.05}&
            3.00&\uncertainty{3.04}{+0.28}{-0.23}&\uncertainty{2.92}{+0.05}{-0.06}&\uncertainty{4.22}{+0.04}{-0.04}&\uncertainty{4.97}{+0.03}{-0.03}
            \\
     -21.15 &5.52&\uncertainty{4.85}{+0.59}{-0.63}&\uncertainty{4.54}{+0.29}{-0.28}&\uncertainty{5.72}{+0.20}{-0.20}&\uncertainty{4.08}{+0.13}{-0.13}&
            5.69&\uncertainty{5.68}{+0.45}{-0.55}&\uncertainty{{5.48}}{+0.12}{-0.12}&\uncertainty{7.09}{+0.08}{-0.08}&\uncertainty{7.00}{+0.06}{-0.06}&
            6.39&\uncertainty{6.42}{+0.47}{-0.38}&\uncertainty{6.17}{+0.07}{-0.08}&\uncertainty{5.56}{+0.05}{-0.05}& \uncertainty{6.48}{+0.04}{-0.04}  
            \\
     -20.60 &7.81&\uncertainty{7.31}{+0.90}{-1.05}&\uncertainty{7.02}{+0.35}{-0.35}&\uncertainty{7.34}{+0.22}{-0.22}&\uncertainty{6.17}{+0.16}{-0.16}&
            8.64&\uncertainty{8.63}{+0.77}{-0.83}&\uncertainty{8.40}{+0.13}{-0.13}&\uncertainty{8.45}{+0.09}{-0.09}&\uncertainty{8.39}{+0.07}{-0.07}&
            9.72&\uncertainty{9.91}{+0.75}{-0.61}&\uncertainty{9.52}{+0.10}{-0.10}&\uncertainty{6.54}{+0.05}{-0.05}& \uncertainty{7.64}{+0.04}{-0.04}
            \\
     -20.05 &10.8&\uncertainty{9.83}{+1.26}{-1.34}&\uncertainty{9.82}{+0.41}{-0.39}&\uncertainty{9.37}{+0.25}{-0.25}&\uncertainty{8.17}{+0.18}{-0.18}&
            10.9&\uncertainty{11.0}{+0.96}{-1.06}&\uncertainty{10.7}{+0.16}{-0.15}&\uncertainty{9.57}{+0.09}{-0.09}&\uncertainty{9.47}{+0.07}{-0.07}&
            12.1&\uncertainty{12.3}{+0.92}{-0.78}&\uncertainty{12.0}{+0.11}{-0.11}&\uncertainty{7.73}{+0.05}{-0.05}&\uncertainty{8.95}{+0.05}{-0.05} 
            \\
     -19.50 &13.5&\uncertainty{12.5}{+1.27}{-1.49}&\uncertainty{12.3}{+0.50}{-0.46}&\uncertainty{11.4}{+0.28}{-0.28}&\uncertainty{10.3}{+0.20}{-0.20}&
            13.3&\uncertainty{13.3}{+1.10}{-1.31}&\uncertainty{13.1}{+0.17}{-0.18}&\uncertainty{11.2}{+0.10}{-0.10}&\uncertainty{11.1}{+0.08}{-0.08}&
            13.9&\uncertainty{14.3}{+0.98}{-0.91}&\uncertainty{14.1}{+0.11}{-0.12}&\uncertainty{9.34}{+0.06}{-0.06}&\uncertainty{10.6}{+0.05}{-0.05}   
            \\
     -18.95 &16.0&\uncertainty{15.5}{+1.75}{-1.90}&\uncertainty{15.3}{+0.52}{-0.52}&\uncertainty{14.1}{+0.31}{-0.31}&\uncertainty{12.7}{+0.22}{-0.22}&
            16.4&\uncertainty{16.6}{+1.22}{-1.40}&\uncertainty{16.3}{+0.20}{-0.20}&\uncertainty{13.9}{+0.11}{-0.11}&\uncertainty{13.7}{+0.09}{-0.09}&
            17.5&\uncertainty{17.5}{+1.06}{-0.88}&\uncertainty{17.7}{+0.17}{-0.17}&\uncertainty{13.5}{+0.09}{-0.09}&\uncertainty{14.5}{+0.08}{-0.08}   
            \\
     -18.40 &20.0&\uncertainty{18.8}{+2.03}{-2.48}&\uncertainty{18.9}{+0.59}{-0.59}&\uncertainty{16.8}{+0.35}{-0.35}&\uncertainty{16.3}{+0.28}{-0.28}&
            20.0&\uncertainty{19.7}{+1.49}{-1.76}&\uncertainty{19.9}{+0.24}{-0.22}&\uncertainty{16.7}{+0.12}{-0.12}&\uncertainty{16.9}{+0.10}{-0.10}&
            20.9&\uncertainty{21.3}{+1.20}{-1.01}&\uncertainty{21.0}{+0.40}{-0.39}&\uncertainty{18.6}{+0.27}{-0.27}&\uncertainty{19.6}{+0.02}{-0.02}   
            \\\cline{12-16}
     -17.85 &22.4&\uncertainty{22.0}{+2.11}{-2.61}&\uncertainty{22.2}{+0.65}{-0.61}&\uncertainty{24.5}{+0.41}{-0.41}&\uncertainty{23.5}{+0.30}{-0.30}&
            22.3&\uncertainty{23.1}{+1.74}{-1.77}&\uncertainty{23.0}{+0.28}{-0.27}&\uncertainty{21.8}{+0.17}{-0.17}&\uncertainty{22.5}{+0.13}{-0.13}&
            10.1&\uncertainty{8.11}{+0.76}{-0.86}&\uncertainty{8.34}{+0.67}{-0.74}&\uncertainty{5.79}{+0.37}{-0.37}&\uncertainty{5.76}{+0.70}{-0.70}   
            \\
     -17.30 &27.9&\uncertainty{27.0}{+2.54}{-2.98}&\uncertainty{27.7}{+0.70}{-0.67}&\uncertainty{36.3}{+0.49}{-0.49}&\uncertainty{32.9}{+0.37}{-0.37}&
            30.1&\uncertainty{28.3}{+2.42}{-2.34}&\uncertainty{30.4}{+0.46}{-0.44}&\uncertainty{28.9}{+0.31}{-0.31}&\uncertainty{28.9}{+0.23}{-0.23}&
            0&0&\uncertainty{0.02}{+0.02}{-0.02}&0&0   
            \\
     -16.75 & 33.8 &\uncertainty{34.5}{+3.32}{-3.52}&\uncertainty{35.1}{+0.79}{-0.79}&\uncertainty{57.6}{+0.62}{-0.62}&\uncertainty{50.1}{+0.50}{-0.50}&
            40.7&\uncertainty{36.5}{+4.34}{-4.29}&\uncertainty{37.7}{+0.92}{-0.92}&\uncertainty{56.7}{+1.02}{-1.02}&\uncertainty{51.4}{+1.66}{-1.66}&
              0&0&0&0&0 
            \\\cline{7-11}
     -16.20 &47.8&\uncertainty{45.3}{+4.06}{+4.71}&\uncertainty{46.9}{+0.85}{-0.87}&\uncertainty{98.2}{+0.97}{-0.97}&\uncertainty{83.4}{+0.73}{-0.73}&
            15.1&\uncertainty{18.9}{+3.72}{-3.71}&\uncertainty{19.7}{+3.14}{-3.20}&\uncertainty{25.7}{+2.12}{-2.12}&\uncertainty{23.2}{+2.13}{-2.13}&
              0&0&0&0&0   
            \\
     -15.65 &59.1&\uncertainty{59.5}{+9.94}{+9.74}&\uncertainty{62.0}{+1.62}{-1.64}&\uncertainty{113}{+1.46}{-1.46}&\uncertainty{93.0}{+1.04}{-1.04}&
              0&0&0&0&0&
              0&0&0&0&0   
              \\
     -15.10 &63.3&\uncertainty{77.8}{+20.4}{-17.5}&\uncertainty{76.5}{+2.45}{-2.31}&\uncertainty{138}{+2.41}{-2.41}&\uncertainty{120}{+1.67}{-1.67}&
              0&0&0&0&0&
              0&0&0&0&0   
              \\
     -14.55 &104&\uncertainty{103}{+30.7}{-28.0}&\uncertainty{117}{+4.77}{-4.77}&\uncertainty{213}{+4.81}{-4.81}&\uncertainty{176}{+3.23}{-3.23}&
              0&0&0&0&0&
              0&0&0&0&0   
              \\
     -14.00 &144&\uncertainty{119}{+34.7}{-33.1}&\uncertainty{145}{+6.62}{-6.77}&  &  &
              0&0&0&0&0&
              0&0&0&0&0   \\
    \hline \cline{2-6}

    \end{tabular}
    \caption{The tabulated luminosity function of the redshift bins $0.05  < z < 0.2$, $0.2 < z < 0.4$, and $0.4 < z < 0.6$, as shown in Fig~\ref{fig:LC_PAUS_W1_W3}. All luminosity function values have been multiplied by a factor of $10^3$ for readability. The horizontal lines indicate the magnitude limits above which the luminosity functions are considered complete using the same methodology as in Fig.~\ref{fig:GALFORM_LF_selection}.}
    \label{tab:LF_Tables_1}
\end{table}
% \end{landscape}

% \begin{landscape}
\begin{table}
    \centering
    \begin{tabular}{cccccccccccccccc}
    % \hline
    %    1& 2& 3& 4& 5& 6& 7& 8& 9&10&11\\
    \hline
         &      \multicolumn{5}{c}{0.6 < $z$ < 0.8} & \multicolumn{5}{c}{0.8 < $z$ < 1.0}\\
    \cmidrule(lr){2-6}  \cmidrule(lr){7-11}
     $M_i$ &$\Phi^{\rm{LC}}_{\rm{true}}$&$\Phi^{\rm{LC}}_{\rm{Gau}}$&$\Phi^{\rm{LC}}_{\rm{BCNZ}}$&$\Phi^{\rm{PAUS}}_{\rm{W1}}$&$\Phi^{\rm{PAUS}}_{\rm{W3}}$&$\Phi^{\rm{LC}}_{\rm{true}}$&$\Phi^{\rm{LC}}_{\rm{Gau}}$&$\Phi^{\rm{LC}}_{\rm{BCNZ}}$&$\Phi^{\rm{PAUS}}_{\rm{W1}}$&$\Phi^{\rm{PAUS}}_{\rm{W3}}$\\
\hline 
     -26.10 & 0&0&0&0&0&
              0&0&0&0&0  
              \\
     -25.55 & 0&0&0&0&0&
              0&0&0&0&0  
              \\
     -25.00 & 0&0&0&0&\uncertainty{0.001}{+0.001}{-0.001}&
              0&0&0&0&\uncertainty{0.002}{+0.001}{-0.001}  
              \\
     -24.45 & 0& 0&\uncertainty{0.001}{+0.001}{-0.001}&\uncertainty{0.002}{+0.001}{-0.001}&\uncertainty{0.007}{+0.001}{-0.001}&
              0& 0&\uncertainty{0.001}{+0.001}{-0.001}&\uncertainty{0.006}{+0.001}{-0.001}&\uncertainty{0.009}{+0.001}{-0.001} 
              \\
     -23.90 & 0& 0&\uncertainty{0.004}{+0.002}{-0.002}&\uncertainty{0.023}{+0.003}{-0.003}&\uncertainty{0.027}{+0.002}{-0.002}&
              0& 0&\uncertainty{0.015}{+0.003}{-0.003}&\uncertainty{0.038}{+0.003}{-0.003}&\uncertainty{0.038}{+0.002}{-0.002}   
              \\
     -23.35 &0.07&\uncertainty{0.09}{+0.01}{-0.01}&\uncertainty{0.08}{+0.01}{-0.01}&\uncertainty{0.19}{+0.01}{-0.01}&\uncertainty{0.172}{+0.005}{-0.005}&
            0.13&\uncertainty{0.13}{+0.01}{-0.01}&\uncertainty{0.15}{+0.01}{-0.01}&\uncertainty{0.191}{+0.006}{-0.006}&  
            \uncertainty{0.162}{+0.005}{-0.005}\\
     -22.80 &0.44&\uncertainty{0.52}{+0.05}{-0.06}&\uncertainty{0.44}{+0.02}{-0.02}&\uncertainty{0.87}{+0.01}{-0.01}&\uncertainty{0.83}{+0.01}{-0.01}&
            0.57&\uncertainty{0.59}{+0.05}{-0.05}&\uncertainty{0.57}{+0.02}{-0.02}&\uncertainty{0.66}{+0.01}{-0.01}&  
            \uncertainty{0.59}{+0.01}{-0.01}\\
     -22.25 &1.21&\uncertainty{1.30}{+0.11}{-0.11}&\uncertainty{1.18}{+0.03}{-0.03}&\uncertainty{2.86}{+0.02}{-0.02}&\uncertainty{2.64}{+0.02}{-0.02}&
            1.32&\uncertainty{1.30}{+0.09}{-0.07}&\uncertainty{1.32}{+0.03}{-0.03}&\uncertainty{1.58}{+0.02}{-0.02}&  
            \uncertainty{1.43}{+0.01}{-0.01}\\
     -21.70 &2.93&\uncertainty{3.08}{+0.21}{-0.21}&\uncertainty{2.87}{+0.04}{-0.04}&\uncertainty{4.91}{+0.03}{-0.03}&\uncertainty{4.17}{+0.02}{-0.02}&
            3.08&\uncertainty{3.06}{+0.17}{-0.14}&\uncertainty{3.11}{+0.04}{-0.05}&\uncertainty{2.53}{+0.02}{-0.02}&  
            \uncertainty{2.27}{+0.02}{-0.02}\\
     -21.15 &6.29&\uncertainty{6.69}{+0.48}{-0.49}&\uncertainty{6.14}{+0.06}{-0.06}&\uncertainty{7.04}{+0.04}{-0.04}&\uncertainty{6.70}{+0.03}{-0.03}&
            6.12&\uncertainty{6.30}{+0.32}{-0.29}&\uncertainty{6.05}{+0.07}{-0.07}&\uncertainty{3.92}{+0.04}{-0.04}&  
            \uncertainty{3.55}{+0.03}{-0.03}\\
     -20.60 &9.78&\uncertainty{10.5}{+0.72}{-0.76}&\uncertainty{9.83}{+0.08}{-0.08}&\uncertainty{8.98}{+0.05}{-0.05}&\uncertainty{8.73}{+0.04}{-0.04}&
            8.28&\uncertainty{8.50}{+0.37}{-0.38}&\uncertainty{7.88}{+0.19}{-0.19}&\uncertainty{5.02}{+0.07}{-0.07}&  
            \uncertainty{4.51}{+0.05}{-0.05}\\\cline{7-11}
     -20.05 &10.4&\uncertainty{11.6}{+0.73}{-0.81}&\uncertainty{11.3}{+0.13}{-0.12}&\uncertainty{10.6}{+0.06}{-0.06}&\uncertainty{10.5}{+0.06}{-0.06}&
            5.46&\uncertainty{5.62}{+0.37}{-0.34}&\uncertainty{5.16}{+0.14}{-0.16}&\uncertainty{4.27}{+0.14}{-0.14}&  
            \uncertainty{4.22}{+0.08}{-0.08}\\\cline{2-6}
     -19.50 &10.0&\uncertainty{11.8}{+0.97}{-1.01}&\uncertainty{11.6}{+0.22}{-0.25}&\uncertainty{11.0}{+0.12}{-0.12}&\uncertainty{11.3}{+0.09}{-0.09}&
            1.69&\uncertainty{1.27}{+0.14}{-0.17}&\uncertainty{1.53}{+0.14}{-0.17}&\uncertainty{0.64}{+0.07}{-0.07}&  
            \uncertainty{0.77}{+0.06}{-0.06}\\
     -18.95 &5.29&\uncertainty{7.05}{+0.67}{-0.72}&\uncertainty{6.50}{+0.33}{-0.35}&\uncertainty{8.18}{+0.35}{-0.35}&\uncertainty{7.97}{+0.21}{-0.21}&
              0.01&0&\uncertainty{0.05}{+0.02}{-0.02}& 0&  
              0\\
     -18.40 & 0& 0&\uncertainty{0.11}{+0.03}{-0.04}&\uncertainty{0.05}{+0.04}{-0.04}&\uncertainty{0.07}{+0.03}{-0.03}&
              0& 0& 0& 0& 0 
              \\ 
    \hline 

    \end{tabular}
    \caption{The same as Table~\ref{tab:LF_Tables_1}, but for the redshift bins $0.6 < z < 0.8$ and $0.8 < z < 1.0$.}
    \label{tab:LF_Tables_2}
\end{table}
\end{landscape}

\end{appendix}

\label{lastpage}
\end{document}